\documentclass[10pt,conference]{IEEEtran}
\usepackage{cite}
\usepackage{amsmath,amssymb,amsfonts}
\usepackage{algorithmic}
\usepackage{graphicx}
\usepackage{textcomp}
\usepackage{xcolor}
\usepackage[hyphens]{url}
\usepackage{fancyhdr}
\usepackage{hyperref}

\newcommand{\hpcayear}{2026}

\title{GRTX: Efficient Ray Tracing for 3D Gaussian-Based Rendering}

\def\hpcacameraready{}

\newcommand\hpcaauthors{Junseo Lee \quad\quad Sangyun Jeon \quad\quad Jungi Lee \quad\quad Junyong Park \quad\quad Jaewoong Sim}
\newcommand\hpcaaffiliation{{} \\ Seoul National University}
\newcommand\hpcaemail{{\{junseo.lee, sangyun.jeon, jungi.lee, junyong.park, jaewoong\}@snu.ac.kr}}

\usepackage{pdfprivacy}
\usepackage{flushend}
\usepackage{xspace}
\usepackage{graphicx}
\usepackage{multirow}
\usepackage{booktabs}
\usepackage[dvipsnames]{xcolor}
\usepackage{caption}
\usepackage{algorithm}
\usepackage{algorithmic}
\usepackage{pifont}
\usepackage{listings}
\usepackage{xhfill}

\newcommand{\putsec}[2]{\section{#2}\label{sec:#1}}
\newcommand{\putssec}[2]{\subsection{#2}\label{ssec:#1}}

\newcommand{\secref}[1]{Section~\ref{sec:#1}}
\newcommand{\ssecref}[1]{Section~\ref{ssec:#1}}

\newcommand{\figref}[1]{Figure~\ref{fig:#1}}
\newcommand{\tabref}[1]{Table~\ref{tab:#1}}
\newcommand{\eqnref}[1]{Equation~\ref{eqn:#1}}

\newcommand{\lstref}[1]{Listing~\ref{lst:#1}}

\let\oldding\ding
\renewcommand{\ding}[2][1]{\scalebox{#1}{\oldding{#2}}}

\definecolor{codegreen}{rgb}{0,0.6,0}
\definecolor{codegray}{rgb}{0.5,0.5,0.5}
\definecolor{codepurple}{rgb}{0.58,0,0.82}
\definecolor{codeblue}{rgb}{0,0,1}
\definecolor{codeorange}{rgb}{0.8,0.4,0}

\lstdefinestyle{shaderstyle}{
    backgroundcolor=\color{white},   
    commentstyle=\color{codeorange},
    keywordstyle=\color{codeblue}\bfseries,
    numberstyle=\tiny\color{codegray},
    stringstyle=\color{codepurple},
    basicstyle=\ttfamily\footnotesize,
    breakatwhitespace=false,         
    breaklines=true,                 
    captionpos=b,                    
    keepspaces=true,                 
    numbers=left,                    
    numbersep=5pt,                  
    showspaces=false,                
    showstringspaces=false,
    showtabs=false,                  
    tabsize=2
}

\lstdefinelanguage{ShaderLang}{
    keywords={if, else, return, void, float, int, bool, vec3, vec4, mat4, while},
    keywords=[2]{rayGenShader, anyHitShader},
    morecomment=[l]{//},
    morestring=[b]",
    sensitive=true,
}

\newcommand{\name}{{GRTX\xspace}}
\newcommand{\textapi}[1]{\texttt{#1}}
\newcommand{\myparagraph}[1]{\vspace{0.02in}\noindent \textbf{#1}}

\author{
  \ifdefined\hpcacameraready
    \IEEEauthorblockN{\hpcaauthors{}}
      \IEEEauthorblockA{
        \hpcaaffiliation{} \\
        \hpcaemail{}
      }
  \else
    \IEEEauthorblockN{\normalsize{HPCA \hpcayear{} Submission
      \textbf{\#\hpcasubmissionnumber{}}} \\
      \IEEEauthorblockA{
        Confidential Draft \\
        Do NOT Distribute!!
      }
    }
  \fi 
}

\fancypagestyle{camerareadyfirstpage}{%
  \fancyhead{}
  
  \fancyhead[C]{
    \ifdefined\aeopen
    \parbox[][12mm][t]{13.5cm}{\hpcayear{} IEEE International Symposium on High-Performance Computer Architecture (HPCA)}    
    \else
      \ifdefined\aereviewed
      \parbox[][12mm][t]{13.5cm}{\hpcayear{} IEEE International Symposium on High-Performance Computer Architecture (HPCA)}
      \else
      \ifdefined\aereproduced
      \parbox[][12mm][t]{13.5cm}{\hpcayear{} IEEE International Symposium on High-Performance Computer Architecture (HPCA)}
      \else
      \parbox[][0mm][t]{13.5cm}{\hpcayear{} IEEE International Symposium on High-Performance Computer Architecture (HPCA)}
    \fi 
    \fi 
    \fi 
    \ifdefined\aeopen 
      \includegraphics[width=12mm,height=12mm]{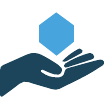}
    \fi 
    \ifdefined\aereviewed
      \includegraphics[width=12mm,height=12mm]{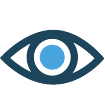}
    \fi 
    \ifdefined\aereproduced
      \includegraphics[width=12mm,height=12mm]{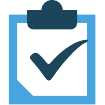}
    \fi
  }
  \fancyfoot[C]{}
}
\begin{document}
\maketitle

\ifdefined\hpcacameraready 
  \thispagestyle{camerareadyfirstpage}
  \pagestyle{empty}
\else
  \thispagestyle{plain}
  \pagestyle{plain}
\fi

\newcommand{\hpcaheight}{0mm}
\ifdefined\eaopen
\renewcommand{\hpcaheight}{12mm}
\fi

\begin{abstract}
  3D Gaussian Splatting has gained widespread adoption across diverse applications due to its exceptional rendering performance and visual quality.
  While most existing methods rely on rasterization to render Gaussians, recent research has started investigating ray tracing approaches to overcome the fundamental limitations inherent in rasterization.
  However, current Gaussian ray tracing methods suffer from inefficiencies such as bloated acceleration structures and redundant node traversals, which greatly degrade ray tracing performance.

  In this work, we present \name{}, a set of software and hardware optimizations that enable efficient ray tracing for 3D Gaussian-based rendering.
  First, we introduce a novel approach for constructing streamlined acceleration structures for Gaussian primitives. Our key insight is that anisotropic Gaussians can be treated as unit spheres through ray space transformations, which substantially reduces BVH size and traversal overhead.
  Second, we propose dedicated hardware support for traversal checkpointing within ray tracing units. This eliminates redundant node visits during multi-round tracing by resuming traversal from checkpointed nodes rather than restarting from the root node in each subsequent round.
  Our evaluation shows that \name{} significantly improves ray tracing performance compared to the baseline ray tracing method with a negligible hardware cost.
\end{abstract}

\putsec{intro}{Introduction}

The recent advent of 3D Gaussian Splatting (3DGS)~\cite{ker:kop23} is rapidly
transforming how we reconstruct and render 3D scenes across a wide range of
applications, including robotics, AR/VR, gaming, and interactive
media~\cite{ye:fu25,pen:sha24,jia:yu24,playcanvas}. 
By representing a scene using a set of anisotropic Gaussians that can be
rendered through rasterization, 3DGS effectively captures fine geometric and
appearance details while generating photorealistic novel views at significantly
higher speeds than prior methods such as NeRF~\cite{mil:sri20}.

While Gaussian primitives can, in principle, be rendered via ray tracing, 3DGS
capitalizes on the computational efficiency of rasterization to achieve
real-time performance.
However, rasterization-based rendering struggles to accurately render scenes
captured with highly distorted cameras~\cite{moe:mir24}---essential for domains
such as robotics and autonomous vehicles---and fails to faithfully reproduce
complex lighting effects that depend on secondary rays, including reflections,
refractions, and shadows.

To address these limitations, recent research from industry-leading companies
such as Google, NVIDIA, and Meta has explored ray tracing for Gaussian
scenes~\cite{moe:mir24,mai:hed25,con:spe24}.
Unfortunately, however, current Gaussian ray tracing methods suffer from
inefficiencies and fall short of 3DGS in terms of performance.
While most methods exploit hardware-accelerated ray-triangle intersection
testing by employing bounding mesh proxies for Gaussian primitives, this
substantially inflates the size of the bounding volume hierarchy (BVH) and
increases memory requirements during BVH traversal.
Furthermore, the multi-round tracing method commonly used in prior work results
in redundant node visits and intersection testing across tracing rounds,
thereby further decreasing traversal efficiency.

In this paper, we present \name{}, a collection of software and hardware
optimizations that greatly improve ray tracing efficiency for Gaussian-based
rendering.
First, we point out that the existing approach to creating acceleration
structures---constructing individual bounding proxy geometries for each
Gaussian and building a single monolithic BVH---is both na\"ive and
inefficient, and that we can actually build acceleration structures that are
far more efficient. 

For this, we exploit a fundamental geometric insight: anisotropic Gaussian
primitives can be uniformly represented as \emph{unit spheres} through ray
space transformations. Importantly, these transformations can be performed
natively by modern ray tracing hardware at the leaf (instance) nodes within the
top-level acceleration structure (TLAS) of a two-level BVH.
Leveraging this insight, we utilize a two-level acceleration structure for
Gaussian ray tracing while constructing only a single, \emph{shared}
bottom-level acceleration structure (BLAS) containing unit sphere geometry. 
All Gaussian primitives then reference the same BLAS in the TLAS, thereby
greatly reducing the BVH memory footprint and traversal overhead compared to
previous ray tracing methods.

Second, we propose enhancing ray tracing hardware with checkpointing and replay
capabilities.
In multi-round ray tracing, BVH nodes intersected by rays in a given round may
fall outside that round's traversal interval, deferring traversal into their
subtrees to subsequent rounds.
To visit their descendants, however, the paths from the root to these nodes
need to be redundantly retraced.
Our key idea is to checkpoint these nodes during the current
round and resume traversal directly from them in the next round, rather than
restarting from the root. This eliminates redundant node visits and
intersection tests while also avoiding unnecessary processing of Gaussians that
have already been found \emph{and} blended in earlier rounds. 

We evaluate \name{} using Vulkan-Sim~\cite{sae:cho22}, a cycle-level graphics
simulator for ray tracing applications, augmented with our in-house ray tracing
simulator that replaces the original ray tracing module.
Our results show that \name{}, combining software and hardware optimizations,
achieves an average speedup of 4.36$\times$ over the baseline employing an
icosahedron bounding mesh.
We also evaluate \name{}-SW, a software-only optimization, on a commodity
GPU, where it achieves speedups of 1.44--2.15$\times$ across the evaluated
scenes.
In summary, this paper makes the following contributions:

\begin{itemize}
\item To our knowledge, this is the first work to identify the challenges and
inefficiencies of rendering Gaussian primitives via ray tracing.
 
\item We present an approach for building efficient acceleration structures
tailored to Gaussian primitives, which leads to substantial reductions in
BVH size and traversal cost.
 
\item We propose checkpointing and replay capabilities for ray tracing
hardware, which help eliminate redundant BVH traversal and intersection
testing during multi-round Gaussian ray tracing.
\end{itemize}

\putsec{back}{Background}

\begin{figure}[t]
  \centering
  \includegraphics[width=0.90\columnwidth]{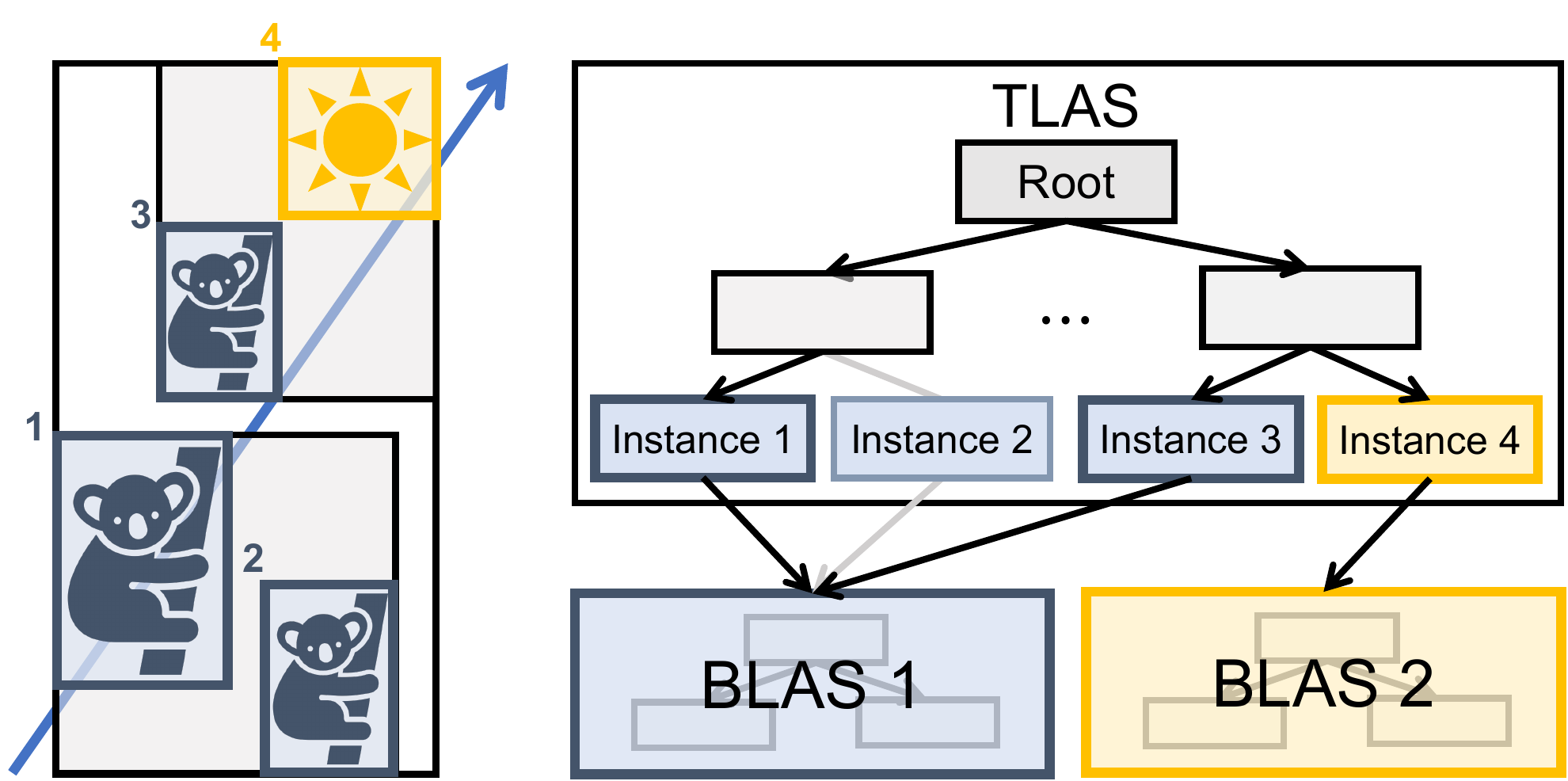}
  \caption{2D visualization of Bounding Volume Hierarchy (BVH) and ray traversal.}
  \vspace{-0.10in}
  \label{fig:bvh-traversal}
\end{figure}

In this section, we first provide background on ray tracing and its
acceleration hardware in modern GPUs. We then briefly introduce the
state-of-the-art method for representing and rendering complex 3D scenes using
a set of Gaussians.

\putssec{back:rt}{Ray Tracing}
Ray tracing is a rendering technique that simulates the path of light from the
viewer to light sources in a 3D scene.
Unlike rasterization-based rendering, which projects scene primitives (e.g.,
triangles) onto the image plane and identifies which pixels they cover, ray
tracing approaches the graphics rendering from the opposite direction---it
begins with rays and determines which primitives these rays intersect.
This allows us to naturally simulate the physical behavior of light, thereby
enabling more accurate rendering of complex optical effects such as
reflections, refractions, and shadows.

\myparagraph{Bounding Volume Hierarchy.}
Since na\"ively testing the intersection between each ray and all the
primitives is prohibitively costly, a spatial data structure called an
acceleration structure (AS) is typically used to reduce the number of
intersection tests.
The AS organizes scene primitives based on their spatial positions, which
allows ray tracers to efficiently skip unnecessary intersection tests.
One of the most widely used acceleration structures is a bounding volume
hierarchy (BVH), a tree-based data structure where each parent node
\emph{spatially} encloses its child nodes. 
This hierarchical relationship provides significant optimizations---when a ray
misses a parent node, all child nodes can be immediately excluded from further
testing.

\figref{bvh-traversal} illustrates an example of ray traversal through a BVH,
where each internal node represents an axis-aligned bounding box (AABB). 
The traversal begins at the root bounding box (Root) and proceeds through the
hierarchy while performing \emph{ray-box} intersection tests to prune
unnecessary branches.
When the ray reaches a leaf node containing scene primitives,
\emph{ray-primitive} intersection tests are conducted and determine the actual
hit point.

In complex scenes with instanced objects, the BVH can be organized as a
two-level hierarchy consisting of a \emph{Top-Level Acceleration Structure}
(TLAS) and \emph{Bottom-Level Acceleration Structures} (BLAS)~\cite{vulkan-as}.
The TLAS serves as the upper level, containing references to BLAS instances as
its leaf nodes. Each BLAS typically represents a distinct object or mesh.
When traversal reaches a TLAS leaf, the ray is transformed into the local
coordinate space of a BLAS instance using a stored transformation matrix, then
traversal continues within that BLAS.
Since the BLAS of a single object can be shared across multiple instances, this
two-level approach significantly reduces memory usage for scenes with repeated
geometry.

\myparagraph{Ray Tracing Accelerators.}
Modern GPUs now feature dedicated hardware accelerators to enhance ray tracing
performance, such as NVIDIA RTX's RT Cores~\cite{rt-core}, AMD RDNA's Ray
Tracing Accelerators~\cite{amd-rdna}, and Intel Arc's Ray Tracing
Units~\cite{intel-rtu}.
While specific implementations and supported features vary across hardware
vendors and architecture generations, these accelerators commonly include
ray-box and ray-triangle intersection test units, thereby alleviating the main
performance bottleneck in ray tracing.
In addition, many accelerators optimize the overall tree traversal process
while autonomously managing traversal stacks, fetching nodes, and handling
ray transformations between TLAS and BLAS.

\begin{figure}[t]
  \centering
  \includegraphics[width=\columnwidth]{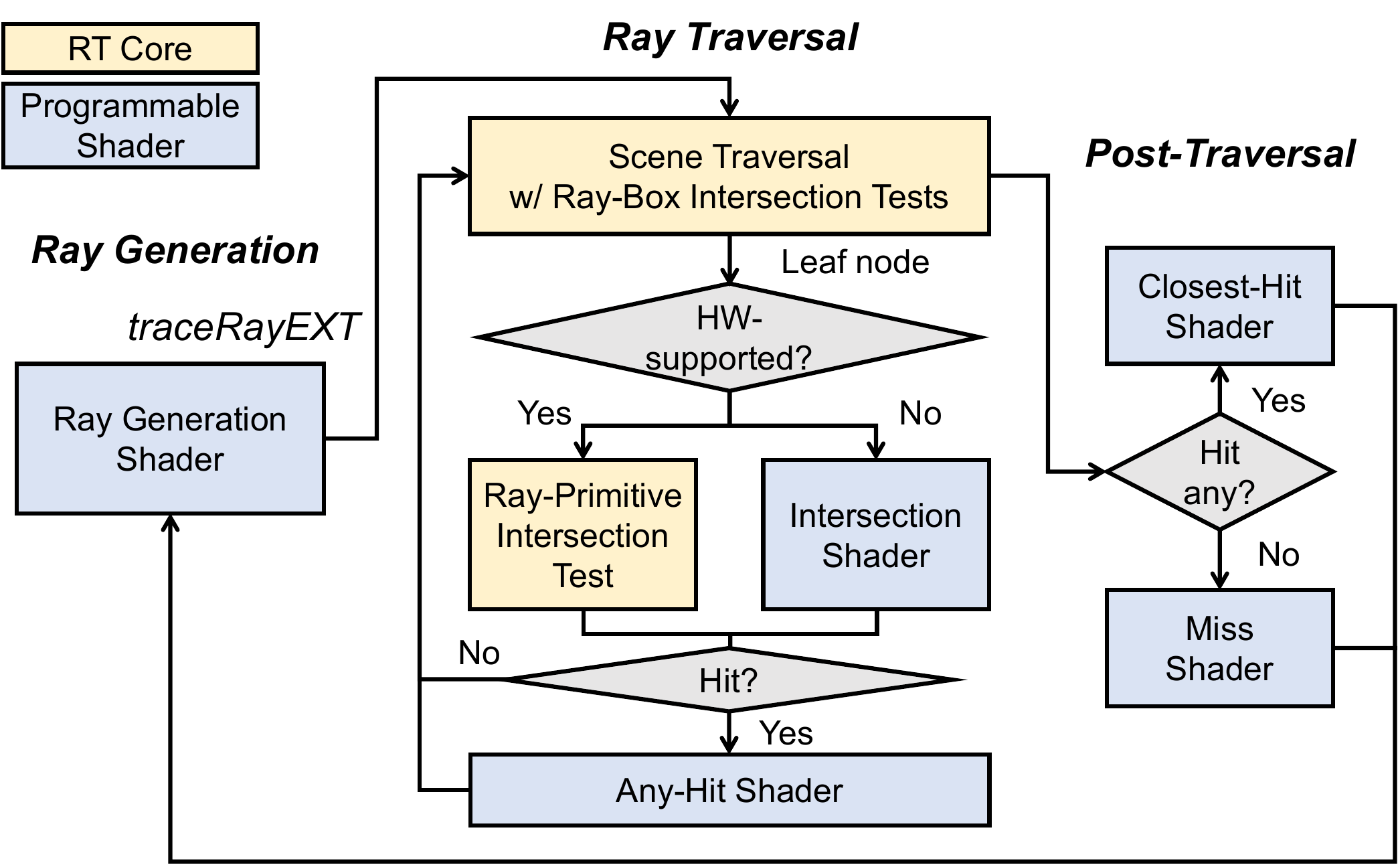}
  \caption{Ray tracing pipeline.}
  \label{fig:rt-pipeline}
\end{figure}

\begin{figure*}[t]
  \centering
  \includegraphics[width=\textwidth]{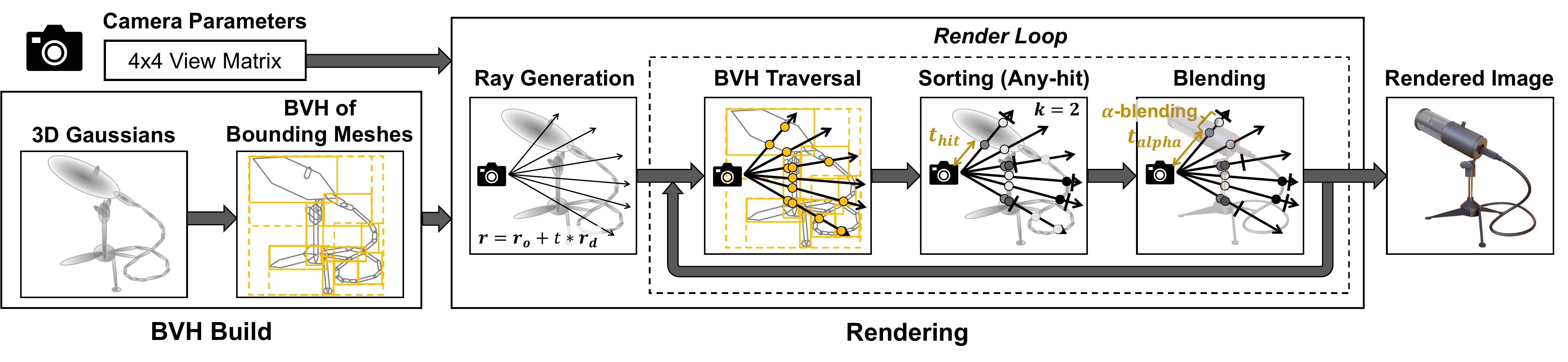}
  \caption{Overview of 3D Gaussian Ray Tracing~\cite{moe:mir24}.}
  \label{fig:3dgrt-pipeline}
\end{figure*}

\myparagraph{Ray Tracing Pipeline.}
\figref{rt-pipeline} shows an overview of the typical ray tracing pipeline
supported by standard graphics APIs such as Vulkan~\cite{vulkan} and
DirectX~\cite{d3d}, or ray tracing frameworks such as OptiX~\cite{optix}.
Conceptually, the pipeline can be divided into three phases: ray generation,
ray traversal, and post-traversal.
Among the operations in the pipeline, shader programs (blue boxes) are executed
on programmable shader cores in GPUs, while tree traversal and intersection
tests (yellow boxes) are usually processed by fixed-function hardware, such
as RT Cores in NVIDIA RTX GPUs.

To begin with, a ray generation shader calls a \textapi{traceRayEXT} function
to initiate the ray tracing process.
This establishes essential parameters for ray tracing, such as ray origin and
direction, along with additional configuration flags.
Once the function is called, the ray begins traversing the BVH to identify
intersected primitives. 
The process continues until either a ray-primitive collision is detected or the
traversal completes.
Note that the traversal can proceed beyond the initial ray-primitive
intersection point when the any-hit shader ignores it via
\textapi{ignoreIntersectionEXT}, allowing evaluation of subsequent potential hits.
For custom primitives, which are not natively supported by hardware,
user-defined shaders can be used for ray-primitive intersection tests.
After the traversal, either the closest-hit shader is invoked (when a hit is
reported) or the miss shader is triggered (when no intersection occurs).
The closest-hit shader can recursively invoke \textapi{traceRayEXT} to cast
secondary rays from the intersection point, enabling effects like reflections,
shadows, and global illumination.

\putssec{back:3dgs}{3D Gaussian-Based Rendering}
3D Gaussian Splatting (3DGS)~\cite{ker:kop23} introduces a novel approach for
representing 3D scenes through anisotropic Gaussian primitives, achieving
state-of-the-art visual quality and rendering performance. 
Each Gaussian is parametrized by spatial properties---its center position
(mean) $\mu$ and 3$\times$3 covariance matrix $\Sigma$---along with visual
attributes including opacity $o$ and spherical harmonic coefficients $sh$ that
encode view-dependent appearance. 
During training, these parameters are optimized to faithfully represent the
scene geometry and appearance. At render time, each Gaussian is treated as an
ellipsoid with defined boundaries around its distribution for computational
efficiency.
There are two methods for rendering Gaussian primitives.

\myparagraph{Method 1: Rasterization.}
3D Gaussian Splatting (3DGS) employs a \emph{rasterization}-based rendering
method~\cite{ker:kop23}.
That is, 3D Gaussians are \emph{projected} onto the image plane as 2D
\emph{splats}, which are sorted by the depth value. 
The final color $\mathbf{C}$ of pixel position $p$ is obtained by accumulating
the colors of overlapping splats through $\alpha$-blending as follows:

\begin{equation}
  \small
  \begin{aligned}
    \mathbf{{C}} = \mathrm{\sum\limits_{i=1}^{N}} \mathrm{\alpha_i\mathbf{c}_i}& \mathrm{\prod_{j=1}^{i-1}} \mathrm{(1-\alpha_j)}, \\
    \textrm{where}~~\mathrm{\alpha_i} = o_\mathrm{i} \times \mathrm{exp}(-\frac{1}{2}&({p}-\mu'_\mathrm{i})^T\Sigma'^{-1}_\mathrm{i}({p}-\mu'_\mathrm{i})).
    \label{eqn:volume-render} 
  \end{aligned}
\end{equation}
Here, $\mathrm{\mathbf{c}_i}$ represents the color, and $\mu'_\mathrm{i}$ and
$\Sigma'_\mathrm{i}$ denote the 2D center and covariance matrix of the i-th
splat, respectively.

\myparagraph{Method 2: Ray Tracing.}
3D Gaussian Ray Tracing (3DGRT) broadens the applicability of 3D Gaussian-based
rendering by addressing fundamental limitations of rasterization. Many recent
works~\cite{moe:mir24,yu:sat24,mai:hed25,gao:gu24,con:spe24} show its potential
by enabling a variety of light effects such as shadow and reflection,
extracting physical properties of a scene, and supporting complex camera
models. In addition, while the original rasterization-based rendering (3DGS)
performs global depth sorting shared across all pixels, ray tracing enables
per-ray sorting that eliminates visual artifacts during camera movement.
However, the advantages of 3DGRT over 3DGS come at a computational cost. The
objective of this work is to mitigate the overhead of Gaussian ray tracing
methods while preserving their benefits.

\putsec{motiv}{Motivation}

In this section, we analyze the Gaussian ray tracer implemented in
3DGRT~\cite{moe:mir24} and compare its rendering performance against 3DGS.
We then identify key inefficiencies in current ray tracing approaches that
motivate the optimizations presented in this work.

\putssec{motiv:3dgrt}{3D Gaussian Ray Tracing}

\figref{3dgrt-pipeline} presents an overview of 3D Gaussian ray
tracing~\cite{moe:mir24}. 
The process begins by constructing a BVH for the scene containing the
Gaussians.
Using this BVH structure along with camera parameters, rendering proceeds
through four key steps: 1) ray generation based on camera parameters, 2) BVH
traversal to identify intersecting Gaussians, 3) depth-based sorting of the
intersecting Gaussians, and 4) alpha blending to compute the final pixel
colors.
In the following, we delve into the three most critical steps in 3DGRT.

\myparagraph{BVH Traversal and Sorting.}
Volume rendering requires accumulating colors from all intersecting Gaussians
in depth order. However, BVH traversal \emph{does not} guarantee that Gaussians
are discovered in this sorted sequence.
This necessitates $N$ traversal rounds over the scene geometry to collect the
next $N$ Gaussians along the ray with a closest-hit shader, which is
computationally expensive.

To address this, 3DGRT employs a $k$-buffer approach~\cite{bav:cal07}, in which
the next $k$ \emph{closest hit} Gaussians are gathered with a \emph{single}
traversal round using an \emph{any-hit} shader.
In ray tracing hardware, the distance ($t_{hit}$) from the camera to each
intersection point is computed during the ray-primitive intersection test.
The any-hit shader then uses the distance value as depth to maintain and update
a $k$-entry buffer, which stores the Gaussians found thus far and keeps them in
depth-sorted order.
Note that the BVH traversal continues through the entire scene until all $k$
closest Gaussians are \emph{definitively} identified; i.e., it visits all the
primitives that will intersect the ray.

The any-hit shader employs distance-based culling to isolate the $k$ closest
Gaussians.
Initially, the traversal interval ($t_{min}$, $t_{max}$) is set to (0,
$\infty$) and is progressively updated throughout each tracing round.
This instructs the RT core to only traverse BVH nodes that intersect the ray
within the range from $t_{min}$ to $t_{max}$. 
The $k$-buffer gradually fills with intersected Gaussians while sorting the
buffer. When the buffer reaches capacity, it performs insertion sort to
maintain only the $k$ closest Gaussians, evicting the farthest one when a
closer Gaussian is encountered.

Any subsequent Gaussian with a $t_{hit}$ value exceeding the largest in the buffer
triggers a hit report rather than being inserted into the buffer.
This report updates $t_{max}$ to the current Gaussian's $t_{hit}$ value,
instructing the RT core to restrict traversal only to the Gaussians and
bounding volumes with smaller $t_{hit}$ values.
Traversal terminates when no candidates remain within the restricted $t_{max}$
threshold. The final $k$-buffer contains the sorted indices and $t_{hit}$
values of exactly the $k$ closest Gaussians, optimally prepared for blending.

\myparagraph{Alpha Blending.}
After the traversal, the colors of the Gaussians in the $k$-buffer are
$\alpha$-blended in order, as shown in~\eqnref{volume-render}.
Instead of pre-computing Gaussian colors as in 3DGS, however, 3DGRT obtains
view-dependent colors per ray using SH coefficients and ray direction at
runtime.
As such, the alpha of the Gaussian is computed using the equation below:

\begin{equation}
  \small
  \alpha = {o} \times G(\mathbf{r_o} + t_{alpha} \mathbf{r_d}) \\ \nonumber
  \text{, where } t_{alpha} = \frac{(\mathbf{\mu} - \mathbf{r_o})^T \Sigma^{-1} \mathbf{r_d}}{\mathbf{r_d}^T \Sigma^{-1} \mathbf{r_d}}. \\ \nonumber
  \label{eqn:t-max-resp}
\end{equation}

Here, $G$ denotes a Gaussian function, with $\mathbf{r_o}$ and $\mathbf{r_d}$
representing the ray origin and direction, respectively. 
The parameter $t_{alpha}$ is the evaluation point for alpha computation,
positioned where the Gaussian achieves maximum response along the ray
trajectory.

After the blending operation, rays can be early terminated to reduce
computational overhead when the accumulated alpha exceeds a predefined
threshold.
For rays that continue, tracing resumes from the $t_{hit}$ value of the last
blended Gaussian (i.e., $t_{min} = max(t_{hit})$) and proceeds until either all rays
terminate or the traversal is complete.

\putssec{motiv:perf}{Performance Analysis}

\begin{figure}[t]
  \centering
  \includegraphics[width=\columnwidth]{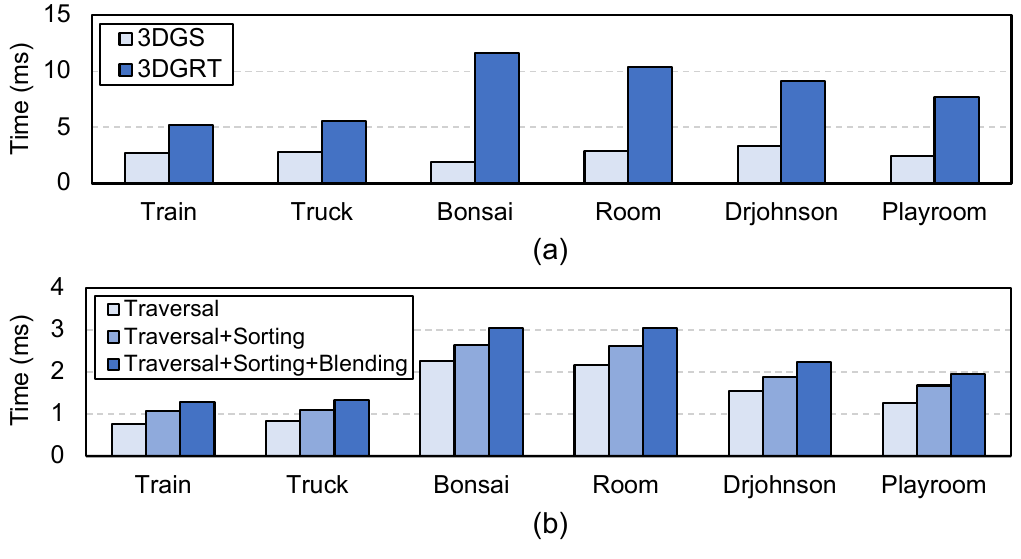}
  \vspace{-0.15in}
  \caption{
    (a) Rendering performance of rasterization (3DGS) and ray tracing (3DGRT). 
    (b) Execution time for a single round of tracing while isolating each operation in 3DGRT.
  }
  \label{fig:3dgs-vs-3dgrt}
  \vspace{-0.20in}
\end{figure}

We present a performance comparison between Gaussian ray
tracing~\cite{moe:mir24} and the original rasterization-based rendering
(i.e., 3D Gaussian Splatting~\cite{ker:kop23}). 
Using the official implementations of both 3DGS and 3DGRT, we train Gaussian
models for 30K iterations and evaluate on several real-world scenes using an
RTX 5090 GPU.
As shown in \figref{3dgs-vs-3dgrt}(a), ray tracing-based Gaussian rendering
is on average approximately 3.04$\times$ slower than rasterization;
note that 3DGS could achieve even higher performance with additional
optimizations.
These results show that a large performance gap remains even with the aid of RT
cores, indicating the need for further optimization.

To identify the performance bottleneck in Gaussian ray tracing, we
incrementally add operations to the ray tracing pipeline and measure the
execution time for a \emph{single} tracing round comprising three operations:
BVH traversal, per-ray sorting in the any-hit shader, and alpha blending in the
raygen shader.
While these operations execute concurrently within a single ray tracing API
call---making precise isolation challenging---we can identify bottlenecks by
observing significant increases in execution time when introducing each operation.
As shown in \figref{3dgs-vs-3dgrt}(b), BVH traversal dominates execution time,
while sorting and blending contribute only marginally.

While 3DGS can directly identify which pixels intersect with Gaussians after 2D
projection, 3DGRT requires exhaustive pointer chasing from root to leaf nodes
for each ray to find intersecting primitives. 
Consequently, the performance gap becomes wider in scenes like Bonsai, where
numerous small Gaussians are concentrated in specific regions, as this
increases traversal time for rays passing through these dense areas. 
Conversely, when Gaussians are distributed more uniformly across the scene, the
performance gap narrows, as observed in scenes like Train and Truck.

\putssec{}{Observations and Opportunities}

\begin{figure}[t]
  \centering
  \includegraphics[width=\columnwidth]{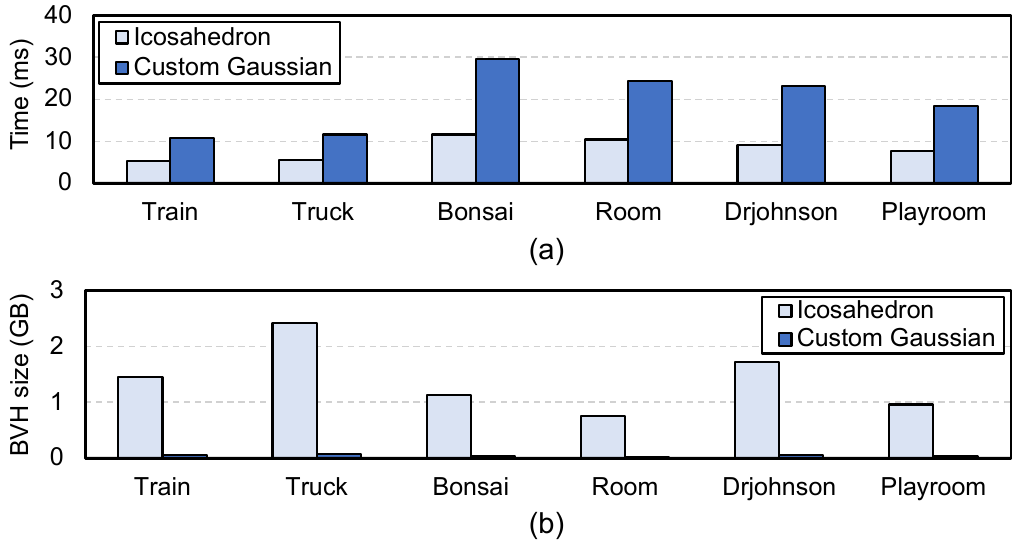}
  \caption{(a) Rendering time and (b) BVH size when using triangles or custom primitives.}
  \label{fig:icosahedron-vs-gaussian}
\end{figure}

\myparagraph{{Observation I: Acceleration Structures and Bounding Primitives.}}
In Gaussian ray tracing, selecting effective bounding primitives for
anisotropic Gaussians is crucial when building an acceleration structure (BVH).
Prior work primarily considers two types of geometric primitives: bounding
triangle meshes~\cite{moe:mir24,con:spe24} or custom Gaussian (ellipsoid)
primitives~\cite{mai:hed25,bla:des25}. 
While rendering quality remains the same regardless of bounding primitives,
each offers distinct advantages and limitations, which lead to the differences
in rendering performance.

\figref{icosahedron-vs-gaussian}(a) compares rendering performance between two
approaches for representing Gaussians in the BVH: a stretched regular
icosahedron (20-faced polyhedron mesh) versus a custom ellipsoid primitive.
Ideally, we want to insert just one primitive per Gaussian into the BVH to
minimize node count and reduce traversal overhead.
While custom primitives enable this one-to-one representation, the
ray-primitive intersection tests need to be performed in software via
user-defined shaders, which are substantially slower than hardware-accelerated
ray-triangle intersection tests.

Using triangle meshes allows us to exploit the ray-triangle intersection
hardware available in modern GPUs, resulting in faster rendering compared to
using custom primitives.
However, reasonably approximating a single Gaussian geometry requires a large
number of triangle primitives, which increases BVH sizes and potentially more
node visits, as shown in \figref{icosahedron-vs-gaussian}(b).
Furthermore, assigning separate bounding primitives to each Gaussian hurts the
cache hit rates, as a scene typically contains hundreds of thousands or
millions of Gaussians. 

Ultimately, we still need a more effective solution to reduce the BVH size and
better utilize the on-chip cache while still leveraging the intersection test
hardware in GPUs.
\ssecref{grtx-sw} introduces our BVH construction strategy for Gaussian ray tracing
before delving into hardware optimization techniques.

\begin{figure}[t]
  \centering
  \includegraphics[width=\columnwidth]{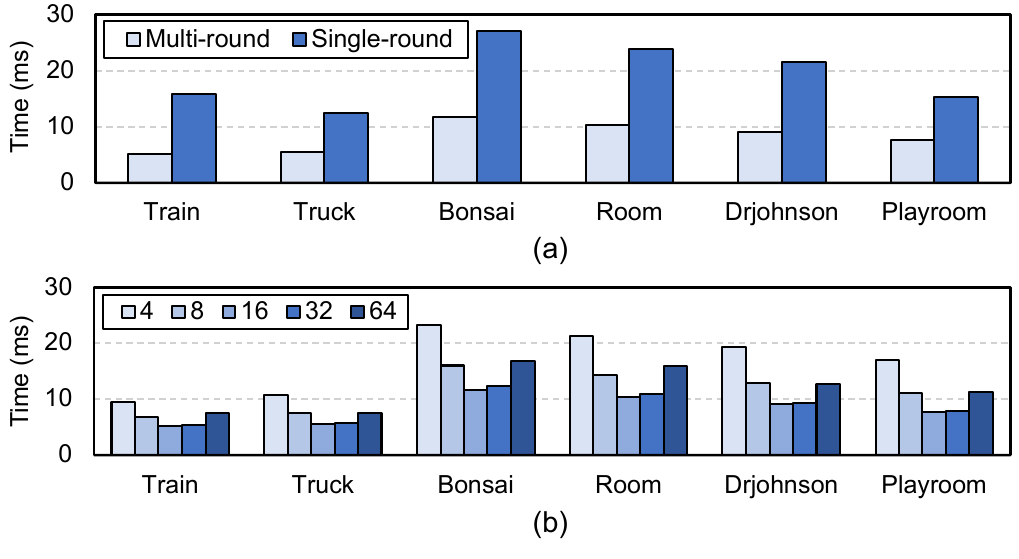}
  \caption{(a) Performance comparison of single-round and multi-round traversal
  methods when $k=16$. (b) Rendering time with different $k$ values.}
  \label{fig:k-sweep}
\end{figure}

\begin{figure}[t]
  \centering
  \includegraphics[width=\columnwidth]{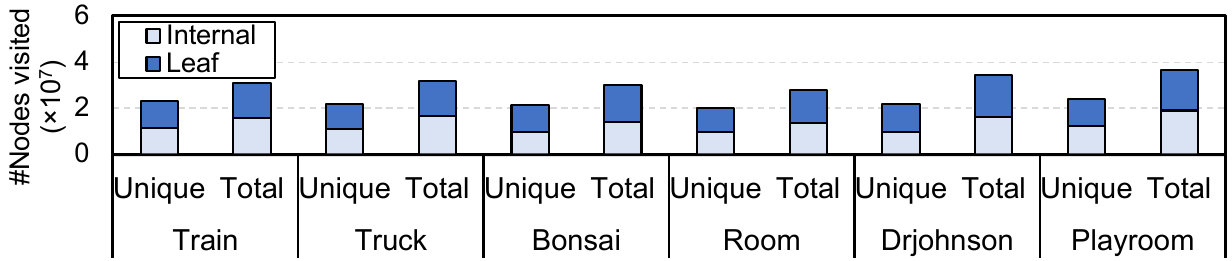}
  \caption{Number of unique versus total visited nodes when $k=16$. The data is
  extracted from Vulkan-Sim.}
  \label{fig:redundant-traversal}
\end{figure}

\myparagraph{Observation II: Redundant BVH Traversal.}
Early ray termination (ERT), a widely used optimization technique in volume
rendering, stops traversal when accumulated alpha exceeds a threshold,
effectively reducing BVH traversal costs.
\figref{k-sweep}(a) shows that multi-round traversal, which collects $k$
Gaussians per round and enables ERT between rounds, outperforms single-round
traversal that collects all intersected Gaussians before blending.
For single-round traversal, we use a large $k$ value (512--2048, depending
on the scene) sufficient to store all intersected Gaussians.
We collect all Gaussians in the any-hit shader without sorting, then perform
sorting and blending after traversal.
The results indicate that multi-round traversal reduces
unnecessary traversal and sorting overhead for Gaussians that will 
not be blended due to ERT.
Although prior work~\cite{mai:hed25,moe:mir24} adopts this multi-round approach
for performance, it suffers from redundancy; the RT core restarts from the root
node each round despite tracing the same ray, leading to repeated node
visits and intersection tests.

The choice of $k$ presents a fundamental trade-off between the benefit of ERT
and the redundancy between multiple BVH traversals.
With a smaller $k$, ERT can be applied in a more fine-grained manner, thereby
reducing the number of unnecessary node accesses and intersection tests during
traversal. However, this requires more redundant traversals of internal
and leaf nodes that have already been visited. 
Conversely, a larger $k$ reduces the number of redundant traversals but
increases unnecessary intersection tests for Gaussians that ultimately do not
contribute to the final pixel color due to early ray termination.
The extreme case of large $k$ is single-round traversal, which eliminates
redundancy but performs poorly due to excessive traversal and sorting
beyond the ERT point.

\figref{redundant-traversal} quantifies this redundancy by showing unique and
total node accesses and intersection tests across multiple rounds when using
$k=16$, which achieves the best performance, as shown in \figref{k-sweep}(b).
We observe that there is a non-negligible gap between unique and total
accesses, which implies that numerous BVH nodes are revisited and tested across
rounds.
Given that traversal constitutes the primary bottleneck in Gaussian ray tracing
and these operations are memory latency-bound, eliminating this redundancy can
effectively improve rendering performance.
In \ssecref{grtx-hw}, we introduce a checkpointing mechanism that enables
subsequent rounds to resume from where previous rounds left off, rather than
restarting from the root node.

\putsec{arch}{\name{}: Gaussian Ray Tracing Acceleration}

In this section, we present \name{}, software and hardware optimization
techniques for Gaussian ray tracing.
Our software optimization aims to accelerate BVH traversal by reducing the BVH
size and its memory footprint while still exploiting hardware-accelerated
ray-primitive intersection.
Our hardware optimization introduces a checkpointing and replay mechanism
to eliminate redundant BVH traversal and intersection testing across multiple
rounds for each ray.

\putssec{grtx-sw}{\name{}-SW: Leveraging Two-Level Acceleration Structure for Gaussian Primitives}

Existing Gaussian ray tracing methods construct a single monolithic BVH for the
entire scene while treating each Gaussian ellipsoid as a \emph{separate}
primitive.
This results in excessively large BVH structures, particularly when
encapsulating Gaussians with bounding meshes (e.g., 20 or 80 triangles per
ellipsoid) to utilize ray-triangle intersection hardware.
For instance, the Truck scene with 2.43M Gaussians requires a BVH size of
approximately 2.42~GB when using 20-triangle bounding meshes.

Instead, we propose leveraging a two-level acceleration structure with a single
\emph{shared} base BLAS across all Gaussian primitives in a scene.
Our key insight is that Gaussian ellipsoids can be treated as \emph{unit
spheres} once rays are transformed into their local coordinate systems, thereby
eliminating the need to build individual BLAS for each Gaussian when utilizing
two-level acceleration.
This greatly reduces the BVH size---down to 432~MB for Truck---and the
memory footprint during BVH traversal while also increasing the cache hit rate.

\begin{figure}[t]
  \centering
  \includegraphics[width=\columnwidth]{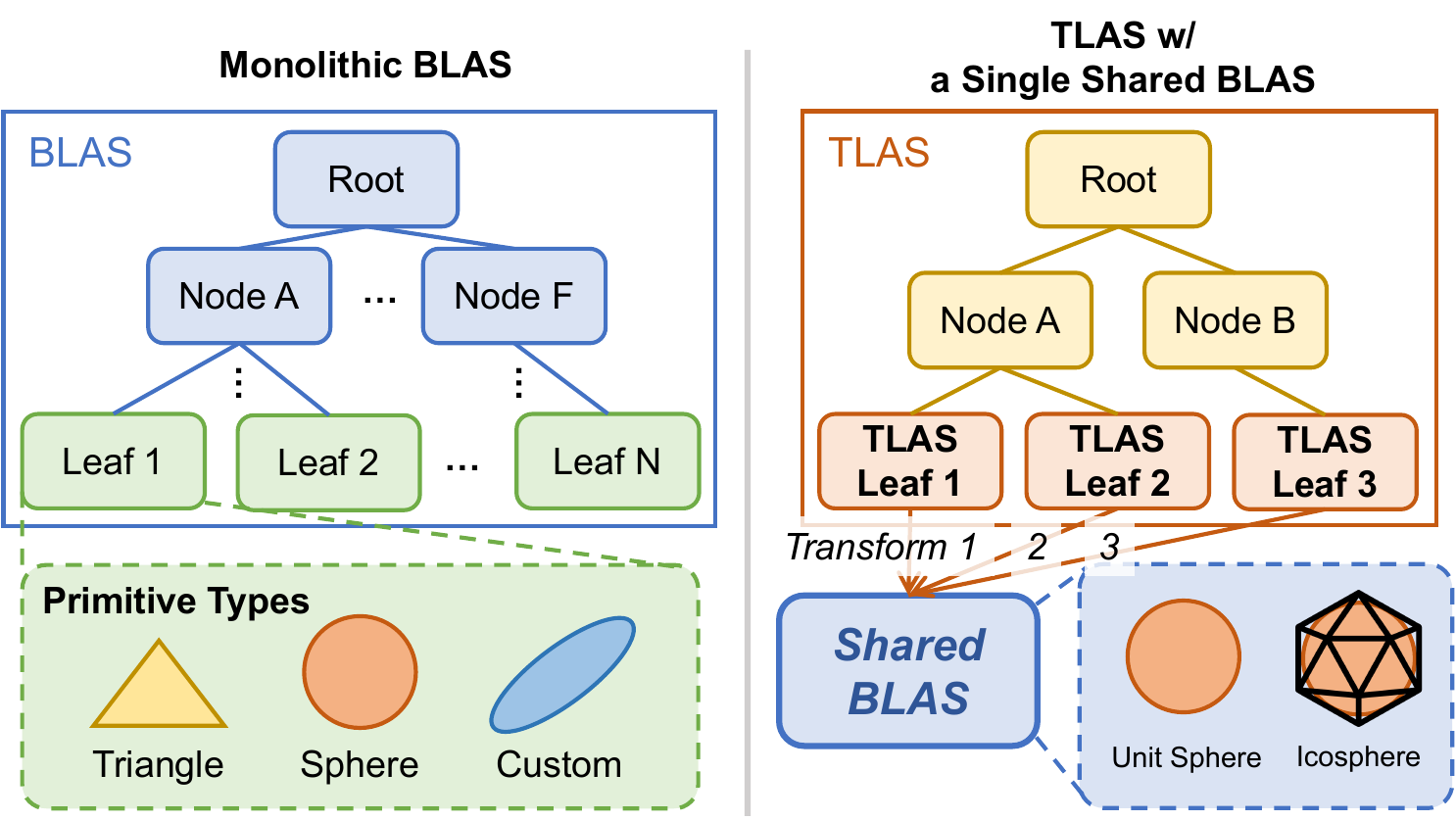}
  \caption{Difference between a monolithic BVH AS and a TLAS with a shared BLAS BVH structure.}
  \label{fig:blas-vs-tlas}
\end{figure}

\figref{blas-vs-tlas} compares a monolithic BVH (i.e., all primitives in a
single BVH) approach used in prior work~\cite{moe:mir24,con:spe24} and our
proposed method of building BVHs for Gaussian ray tracing. 
In two-level acceleration, when a ray hits a leaf node in the TLAS, it is
transformed to the local coordinate system by the transform matrix of each
Gaussian, which is derived from its rotation and scaling matrices.
Modern ray tracing hardware provides native support for this instance
transform~\cite{amd-rdna, rtx5090}.
Then, either a ray-sphere intersection test is performed when using a unit
sphere as a primitive, or additional BLAS traversal occurs when using triangles
as primitives.

\myparagraph{Bounding Primitives for Gaussians.}
To exploit ray-primitive intersection hardware, we consider two alternative
methods: 1) using a unit sphere, or 2) using an icosphere with multiple
triangles.
First, we can directly use a unit sphere as a primitive, which is optimal in
terms of minimizing false positive intersection tests.
After ray transformation, the Gaussian ellipsoid is equivalent to a unit
sphere, so the sphere primitive exactly matches the Gaussian geometry.
This avoids false positive intersections, which are the cases where a
ray intersects the bounding primitive but not the actual Gaussian.
In recent GPU architectures like NVIDIA Blackwell, RT cores natively
support ray-sphere intersection tests that can be performed in hardware.
This requires only one ray-AABB and one ray-sphere intersection test to
determine whether a ray hits a Gaussian after transformation at the TLAS
leaf node.

Second, instead of using a unit sphere, we can use an icosphere with multiple
triangles.
This method is similar to previous approaches that use a stretched icosahedron
mesh~\cite{moe:mir24} and a high-poly icosphere~\cite{con:spe24} to
approximate Gaussian geometry.
The key advantage of this method compared to the first one is that it can
exploit high-throughput ray-triangle intersection test hardware in general RT
units.
While this approach may incur false positives in the intersection test, these
can be mitigated by using a larger number of triangles (e.g., 80 triangles), as
discussed by Condor et al.~\cite{con:spe24}.
However, unlike their monolithic BVH, where the number of leaf nodes scales with
the triangle count per mesh---resulting in multi-gigabyte BVH sizes---our
shared BLAS keeps the overall BVH size small by storing only one template mesh
of a few kilobytes.

We provide quantitative comparisons of these two approaches in
Sections~\ref{ssec:perf} and \ref{sec:analysis}.

\begin{figure*}[t]
  \centering
  \includegraphics[width=\linewidth]{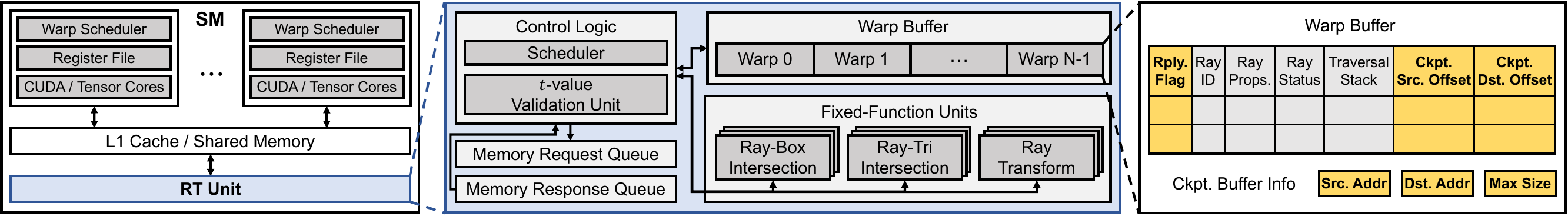}
  \caption{Baseline GPU architecture modeled in Vulkan-Sim~\cite{sae:cho22} with
  \name{}. Extended hardware is highlighted in yellow.}
  \label{fig:arch}
\end{figure*}

\putssec{grtx-hw}{\name{}-HW: HW Acceleration for Gaussian Ray Tracing}

\begin{figure}[t]
  \centering
  \includegraphics[width=\columnwidth]{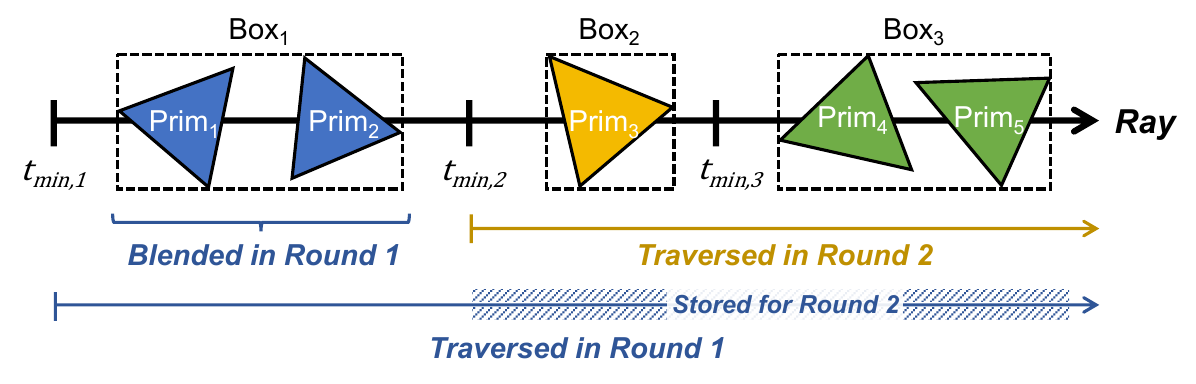}
  \caption{High-level overview of checkpointing and replay mechanism in \name{}-HW.}
  \label{fig:ckpt-rply-overview}
\end{figure}

\myparagraph{Baseline Architecture and Operations.}
\figref{arch} presents the baseline GPU architecture modeled in
Vulkan-Sim~\cite{sae:cho22} along with \name{} hardware extensions for ray
tracing (RT) units.
Each streaming multiprocessor (SM) contains a single RT unit comprising a
dedicated scheduler, a warp buffer, and three types of fixed-function units:
ray-box intersection units, ray-triangle intersection units, and ray
transformation units.

Upon invoking the \texttt{traceRayEXT} intrinsic, a warp delegates its BVH
traversal to the RT unit, which processes the ray information for each thread
in the warp.
The warp buffer stores and manages the associated per-ray data, including ray
status (e.g., active or terminated), ray properties (e.g., ray origin, ray
direction, $t_{min}$, $t_{max}$), and the traversal stack.

At each cycle, the RT scheduler selects a warp to process. 
The RT unit retrieves ray information from the warp buffer and fetches the node
at the top of the traversal stack from memory.
When the node data arrives, the RT unit performs either ray-box or
ray-primitive intersection tests based on the node type.

A hit is reported when two conditions are satisfied: 1) the ray intersects with
the node and 2) the hit point $t_{hit}$ falls within the valid range ($t_{min}
< t_{hit} \le t_{max}$).
Upon detecting a hit, the RT unit pushes the child node address onto the stack.
For TLAS leaf nodes, the RT unit transforms the ray using the transform matrix
stored in the node and pushes the address of the BLAS root node onto the stack.
For the leaf nodes in the BLAS, hit information such as $t_{hit}$ and
primitive ID is also recorded in the warp buffer.

During traversal, when all active rays in a warp hit a primitive or
a timeout occurs, the RT unit invokes the any-hit shader.
When all threads in the warp finish the traversal, the warp retires from
the RT unit and executes the rest of the shader program in the SM.

\myparagraph{Traversal Checkpointing and Replay.}
\figref{ckpt-rply-overview} presents the core concept of our checkpointing
mechanism in \name{}-HW.
During multi-round traversal, each round traces an identical ray but with
different intervals; i.e., $t \in (t_{min,i}, \infty)$ for the $i$-th round.
Because these intervals exhibit substantial overlaps, initiating each traversal
from the root node results in redundant node accesses across rounds.
Our key idea is to checkpoint the nodes and primitives that intersect within
the overlapping intervals between consecutive rounds.
These checkpointed nodes then serve as traversal starting points for the
subsequent round, eliminating the need to retraverse from the root node.
This approach substantially reduces the search space for each round by
constraining traversal to the subtrees rooted at checkpointed nodes, which are
traversed sequentially.

We checkpoint two distinct categories of elements (nodes and primitives).
The first category includes BVH nodes that intersect the ray but are reported
as \emph{missed} because they lie beyond the $k$ closest Gaussians,
making further traversal unnecessary in the current round.
These nodes are identified when they fail the $t_{max}$ test ($t_{hit} >
t_{max}$), and the RT unit stores the nodes in a checkpoint buffer.
The second category comprises \emph{primitives} that intersect the ray and are
reported as \emph{hit} ($t_{hit} \le t_{max}$)---thus invoking the any-hit
shader---but are ultimately rejected because they do not rank among the next
$k$ closest Gaussians for this round.
These rejected primitives are stored in an eviction buffer by the any-hit
shader.

\begin{lstlisting}[caption={Pseudo-code for the any-hit shader and raygen shader.},label={lst:eviction-buffer}]
rayGenShader(){
  while (pixel.alpha < alphaThreshold) {
    moveEvictToKBuf(evictBuffer, prd.evictOffset, kBuffer, k)
    traceRayEXT()
    blendGaussians(pixel, kBuffer)
    if (prd.size < k) break
  }
}

anyHitShader(){
  rejected = insertionSort(kBuffer, tHit, primID)

  if (prd.size == k) {
    evictBuffer[prd.evictOffset] = rejected
    prd.evictOffset++
  }
  prd.size = min(prd.size + 1, k)
  if (tHit < rejected.tHit) {
    ignoreIntersectionEXT()
  }
}
\end{lstlisting}

\lstref{eviction-buffer} shows the pseudo-code for the any-hit and raygen
shaders with the eviction buffer management.
The any-hit shader first stores the rejected Gaussians from the $k$-buffer into
the eviction buffer using the offset stored in the payload (Lines 13-16).
Before the next traversal round starts, the raygen shader sorts and moves the
first $k$ Gaussians from the eviction buffer to the $k$-buffer (Line 3).

\myparagraph{Checkpoint and Eviction Buffer.}
The checkpoint buffer is divided into two types: a source buffer and a
destination buffer.
In each tracing round, traversal starts from the root (first round; replay
flag=0) or resumes from checkpointed nodes in the source buffer (subsequent
rounds; replay flag=1).
During traversal, newly encountered nodes requiring checkpointing
are written to the destination buffer.
To maintain proper buffer indexing, the source and destination offsets in
the warp buffer increment during each read and write to the corresponding
checkpoint buffer.
After each round, the destination buffer becomes the source buffer for the next
round, creating a ping-pong buffer arrangement.

The checkpoint buffer and the eviction buffer require different entry formats
due to their distinct purposes.
Each checkpoint buffer entry contains the node address (8 bytes), the TLAS leaf
node address (8 bytes) if it is a BLAS node, and the corresponding $t_{hit}$
value (4 bytes), totaling 20 bytes per entry.
The TLAS leaf node address is required for correct ray transformation: since we
directly start traversal from a BLAS node (not from TLAS to BLAS), we need to
transform the ray from world space to the object space of each Gaussian using
the transformation matrix stored in the TLAS leaf node.

In contrast, eviction buffer entries have a simpler structure, containing only
the primitive ID (4 bytes) and its $t_{hit}$ value (4 bytes), since the entries
in the eviction buffer will be directly moved to the $k$-buffer in the
subsequent round, where they receive a second opportunity to be accepted as the
$k$ closest Gaussians.
Note that both checkpoint and eviction buffers reside in global memory, not in
the warp buffer, thus they do not require additional storage overhead.

\begin{figure}[t]
  \centering
  \includegraphics[width=\columnwidth]{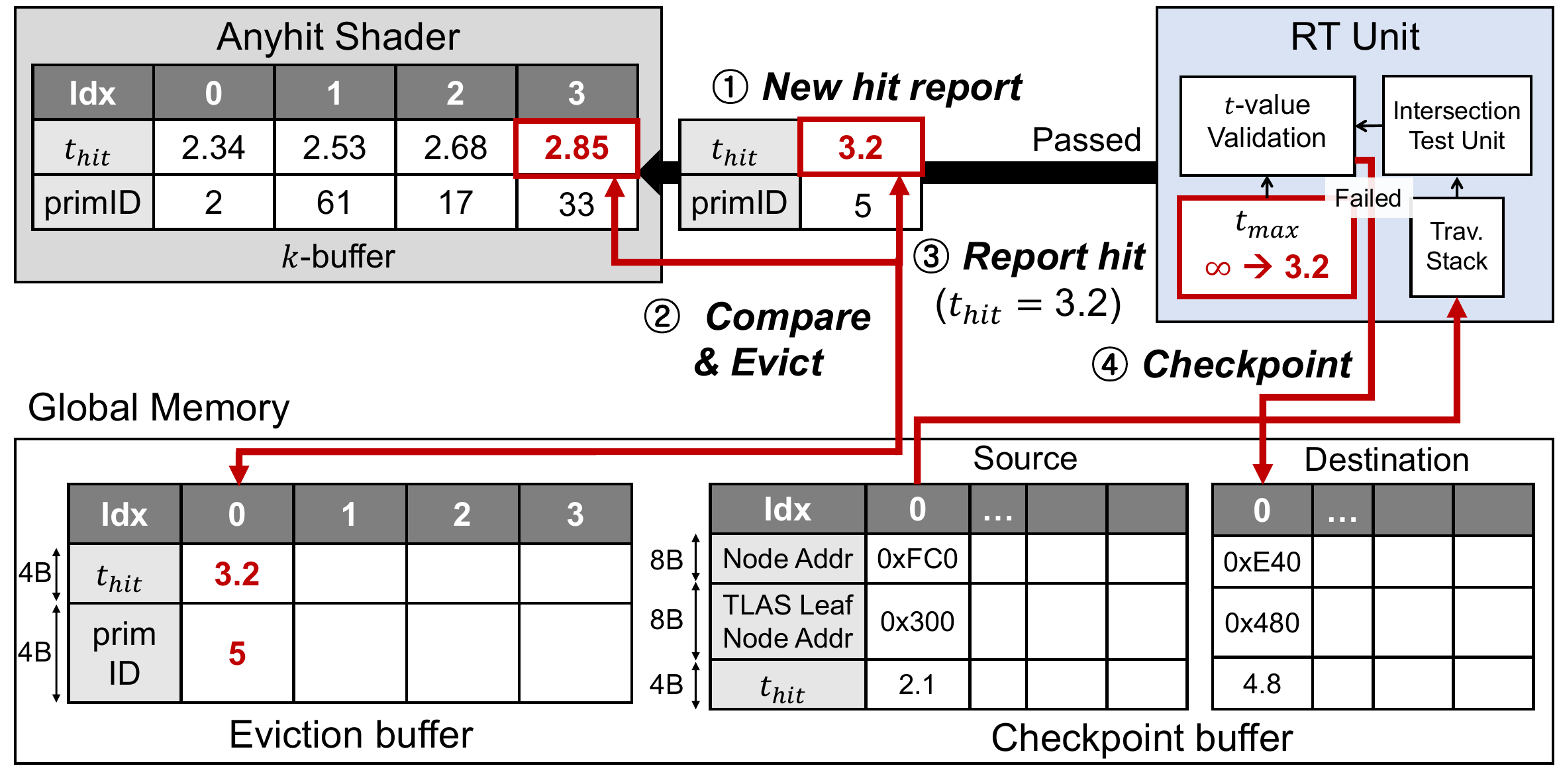}
  \caption{Execution flow of checkpointing and replay.}
  \label{fig:ckpt-rply}
\end{figure}

\myparagraph{Walkthrough Example.}
\figref{ckpt-rply} illustrates the complete execution flow of our checkpointing
and replay mechanism through a concrete example.
Before each round begins, we transfer evicted Gaussians from the eviction
buffer to the $k$-buffer.
During traversal, when the eviction buffer contains more than $k$ Gaussians, we
retain only the $k$ closest to maintain the $k$-buffer size constraint.

Consider a scenario where the $k$-buffer (with $k=4$) is full and encounters a
new hit with $t_{hit} = 3.2$ and primitive ID 5.
The any-hit shader first compares this new hit against the last entry
(i.e., largest $t_{hit}$) of the $k$-buffer
(\raisebox{-0.2ex}{\ding[1.1]{172}}).
Since the new hit is more distant than the current farthest Gaussian in the
$k$-buffer (\texttt{primID} 33 with $t_{hit} = 2.85$), the new primitive (ID 5)
is rejected and is stored in the eviction buffer
(\raisebox{-0.2ex}{\ding[1.1]{173}}).
The shader then reports the hit to the RT unit with $t_{hit} = 3.2$
(\raisebox{-0.2ex}{\ding[1.1]{174}}), which triggers an update of $t_{max}$
from $\infty$ to 3.2.
This updated $t_{max}$ value influences subsequent traversal; the RT unit's
$t$-value validation unit now rejects any intersections beyond this distance,
and nodes failing this test are checkpointed to the destination buffer
(\raisebox{-0.2ex}{\ding[1.1]{175}}).

\putsec{eval}{Evaluation}
\putssec{method}{Methodology}

\begin{table}[t]
  \centering
  \caption{Simulation configuration.}
  \label{tab:sys-config}
  \resizebox{\columnwidth}{!}{%
    {\fontsize{12}{14}\selectfont
      \begin{tabular}{c|c}
        \toprule
        \multicolumn{2}{c}{\textbf{GPU}} \\
        \midrule
        \# Streaming Multiprocessors (SM) & 8, 1365 MHz, in-order  \\
        SIMT Lanes per SM                 & 128 (4 warp schedulers)\\
        L1I Cache                         & 128 KB, 128B line, 16-way LRU, 20 cycles \\
        L1D Cache                         & 128 KB, 128B line, 256-way LRU, 20 cycles \\
        L2  Cache (Unified)               & 4 MB, 128B line, 16-way LRU, 165 cycles  \\
        Memory Clock  & 3500 MHz \\
        \midrule
        \multicolumn{2}{c}{\textbf{Ray Tracing Unit}} \\
        \midrule
        \# RT Units per SM & 1 \\
        Warp Buffer Size   & 8  \\
        \bottomrule
      \end{tabular}
    }
  }
\end{table}


\noindent\textbf{Simulation Infrastructure.}
To evaluate the rendering performance of \name{}, we use
Vulkan-Sim~\cite{sae:cho22}, a cycle-level graphics simulator that runs ray
tracing applications, alongside an in-house cycle-level simulator that models
the ray tracing behavior with any-hit shaders.
The original Vulkan-Sim ray tracing implementation employs a delayed execution
model that completes all ray traversal operations before executing intersection
and any-hit shaders.
However, this does not accurately reflect our baseline GPU behavior, where
any-hit shaders are invoked during traversal, and subsequent traversal is
influenced by any-hit shader results.
For Gaussian ray tracing, traversal can also be early-terminated once the $k$
closest Gaussians are found.
Thus, we develop an in-house ray tracing simulator supporting immediate
shading, which enables any-hit shaders to execute whenever rays in a warp
detect intersected Gaussians, rather than waiting until all traversals
complete. We integrate this RT simulator with Vulkan-Sim.

\tabref{sys-config} shows the GPU configuration used in this work. We use 8 SMs
and scale other parameters based on the RTX 5090 GPU, from which we collected
our motivational data.
We construct the BVH structure using Intel Embree~\cite{wal:woo14},
specifically employing a BVH-6 configuration that supports up to six children
per node.
We compare \name{} against the baseline RT execution that uses a stretched
icosahedral mesh to approximate Gaussian geometry, as in
3DGRT~\cite{moe:mir24}.

We observe that baseline L1 cache hit rates on real GPUs are higher than
those in our simulator, likely due to undisclosed optimizations. 
We assume that GPUs may employ optimizations that lead to an increase in
cache hit rates during BVH traversal. 
We capture this effect by incorporating node prefetching into our simulator: 
upon the first demand fetch of any child leaf node, we issue a one-time prefetch for
its sibling nodes whose bounding boxes are also intersected.
This brings simulated L1 hit rates into closer alignment with those observed on
real hardware.


\begin{table*}[t]
  \centering
  \caption{Summary of workloads. Images are rendered by our Vulkan
  implementation of Gaussian ray tracing. BVH sizes and memory footprints
  are measured using our simulator.}
  \label{tab:workloads}
  \setlength{\tabcolsep}{6pt}
  \resizebox{\linewidth}{!}{%
    \begin{tabular}{cccccccc}
      \toprule
      \multicolumn{2}{c}{\textbf{Dataset}} &
      \multicolumn{2}{c}{\textbf{Tanks\&Temples}~\cite{kna:par17}} &
      \multicolumn{2}{c}{\textbf{Mip-NeRF 360}~\cite{bar:mil22}} &
      \multicolumn{2}{c}{\textbf{Deep Blending}~\cite{hed:phi18}} \\
      \cmidrule(lr){1-2} \cmidrule(lr){3-4} \cmidrule(lr){5-6}
      \cmidrule(lr){7-8}
      \multicolumn{2}{c}{\textbf{Scene}} & \textbf{Train} & \textbf{Truck} &\textbf{Bonsai} &
      \textbf{Room} & \textbf{Drjohnson} & \textbf{Playroom} \\
      & & \includegraphics[width=0.15\linewidth,height=1.8cm]{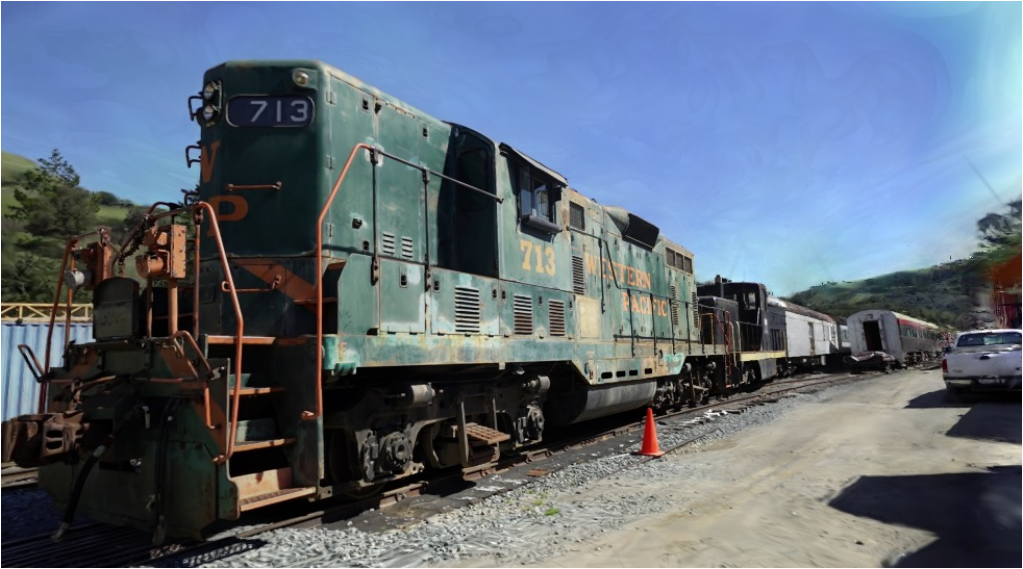} &
      \includegraphics[width=0.15\linewidth,height=1.8cm]{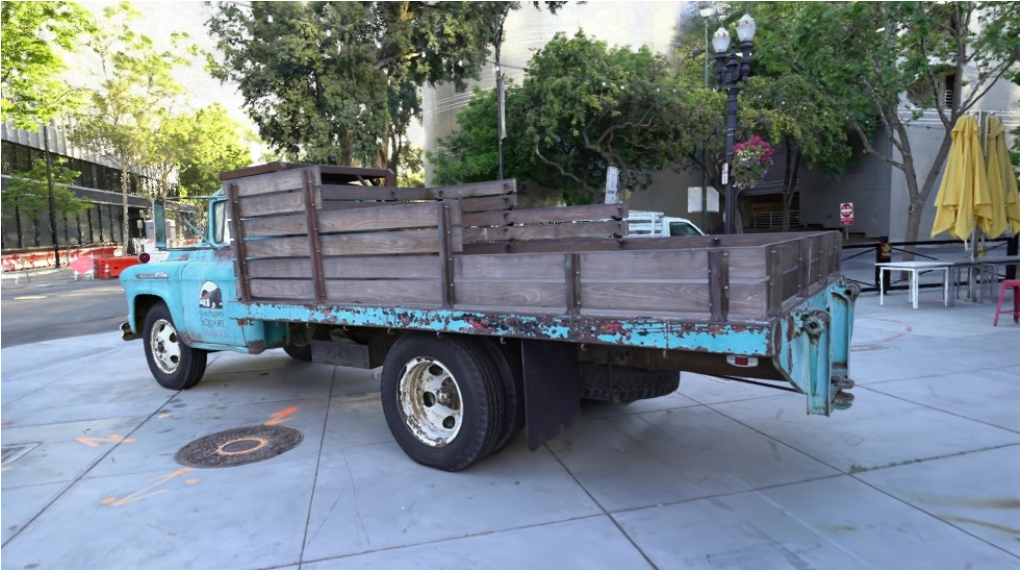} &
      \includegraphics[width=0.15\linewidth,height=1.8cm]{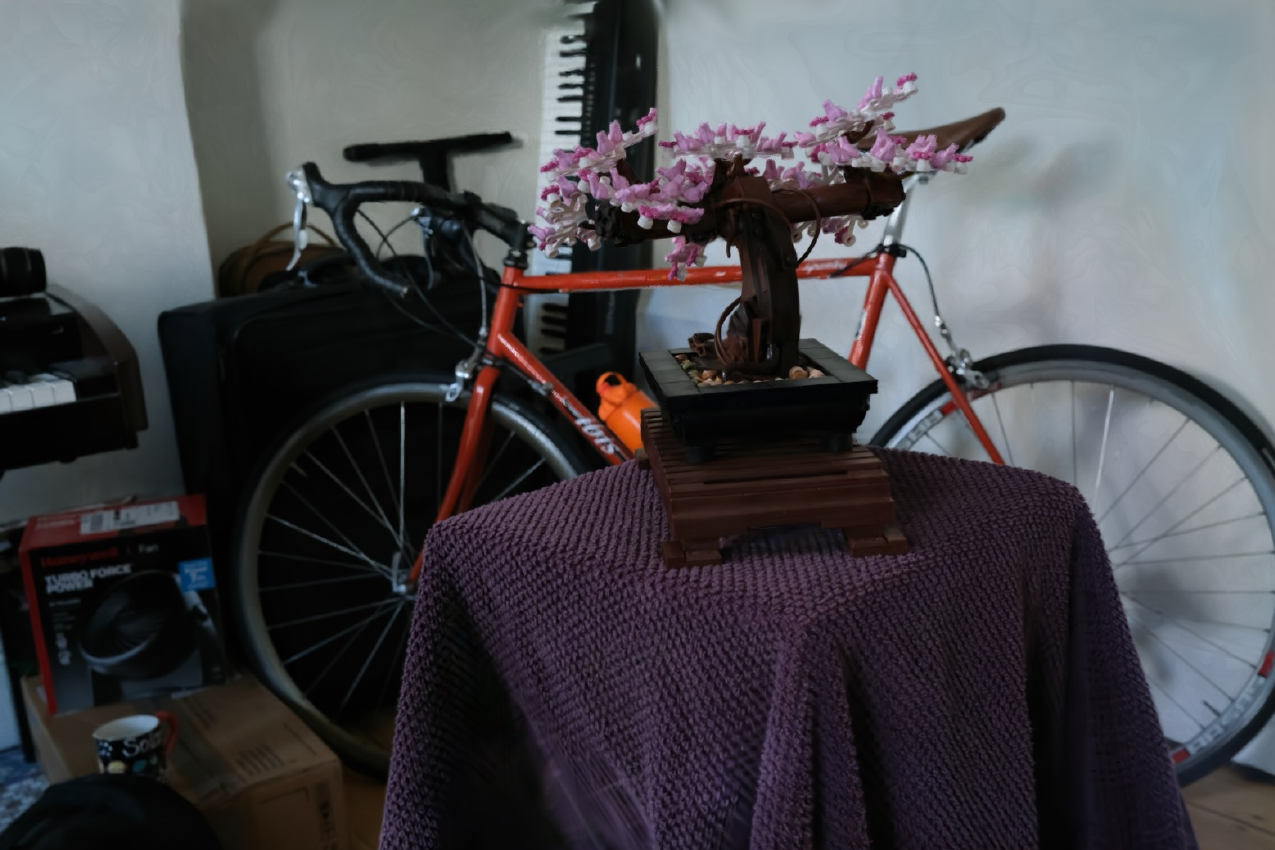} &
      \includegraphics[width=0.15\linewidth,height=1.8cm]{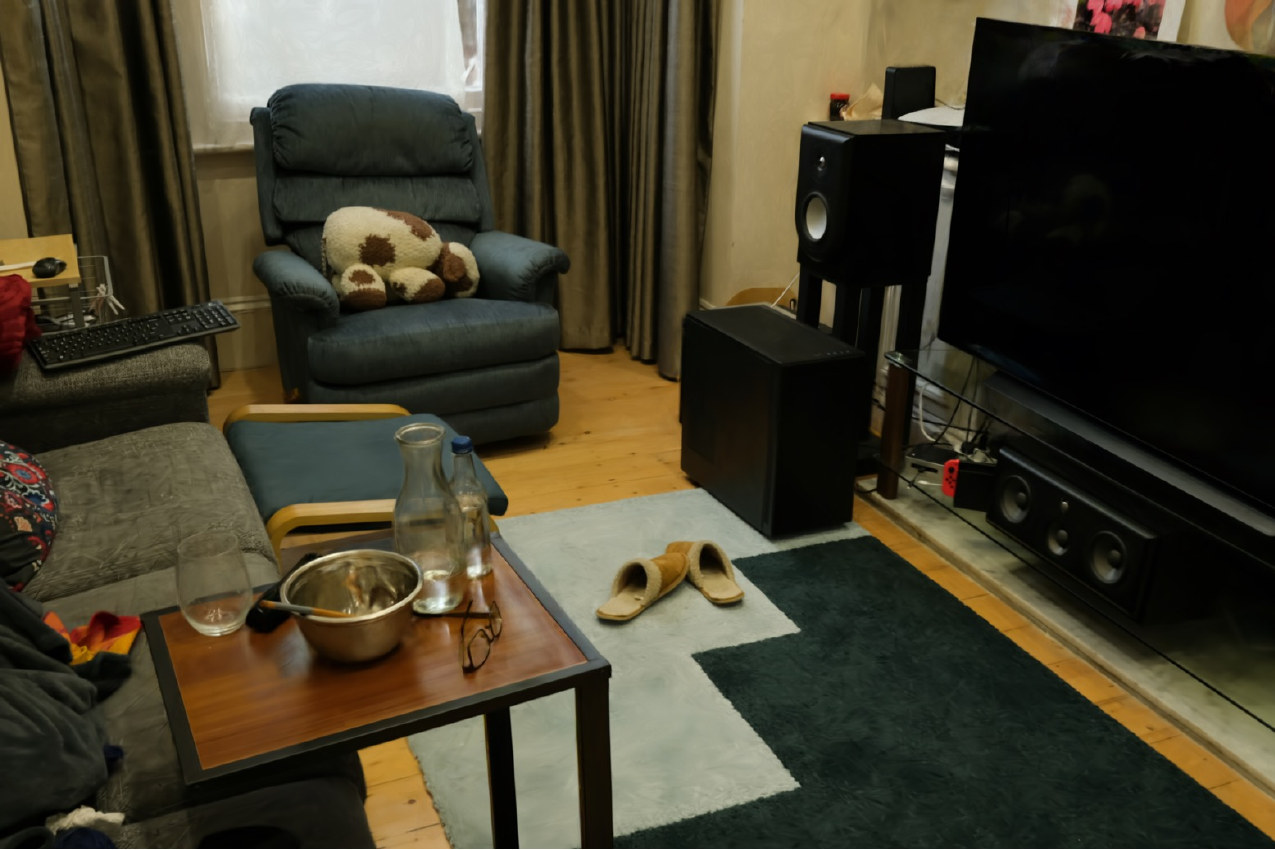} &
      \includegraphics[width=0.15\linewidth,height=1.8cm]{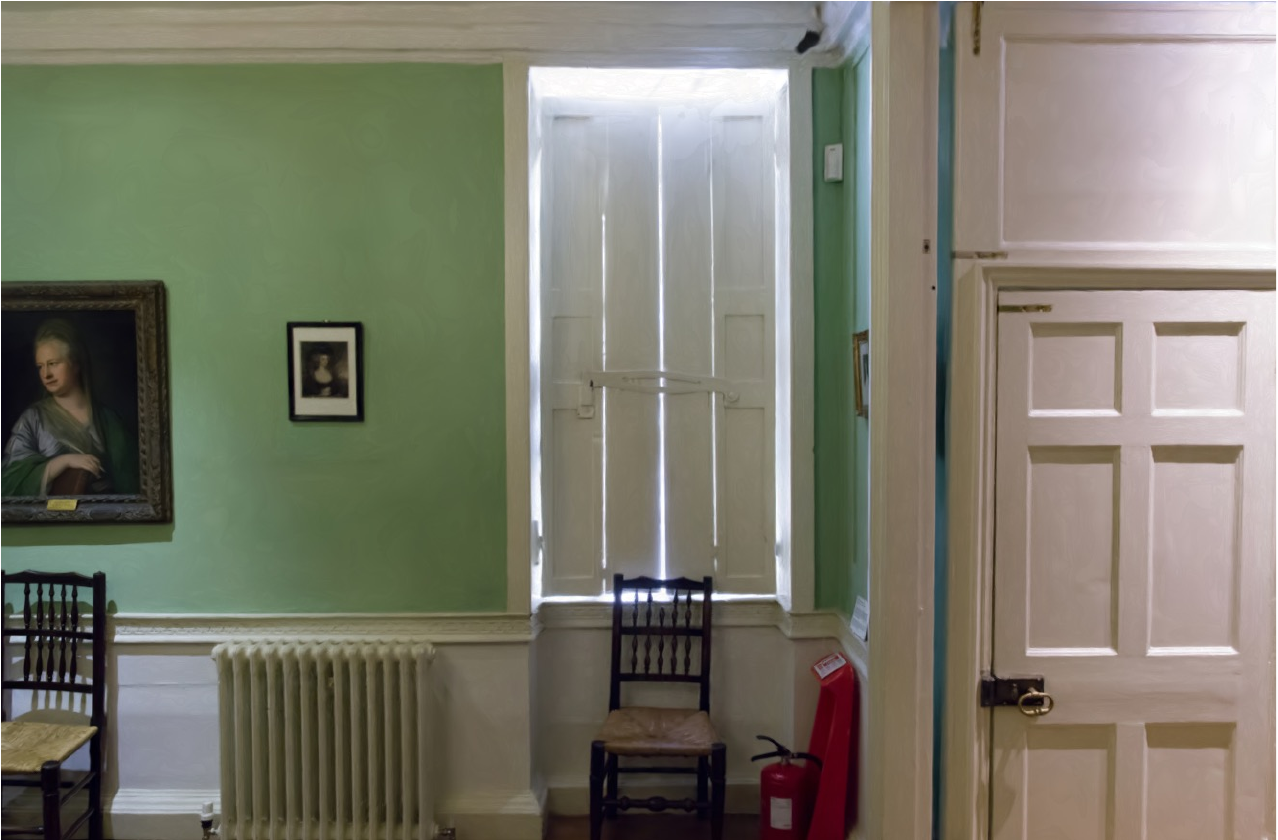} &
      \includegraphics[width=0.15\linewidth,height=1.8cm]{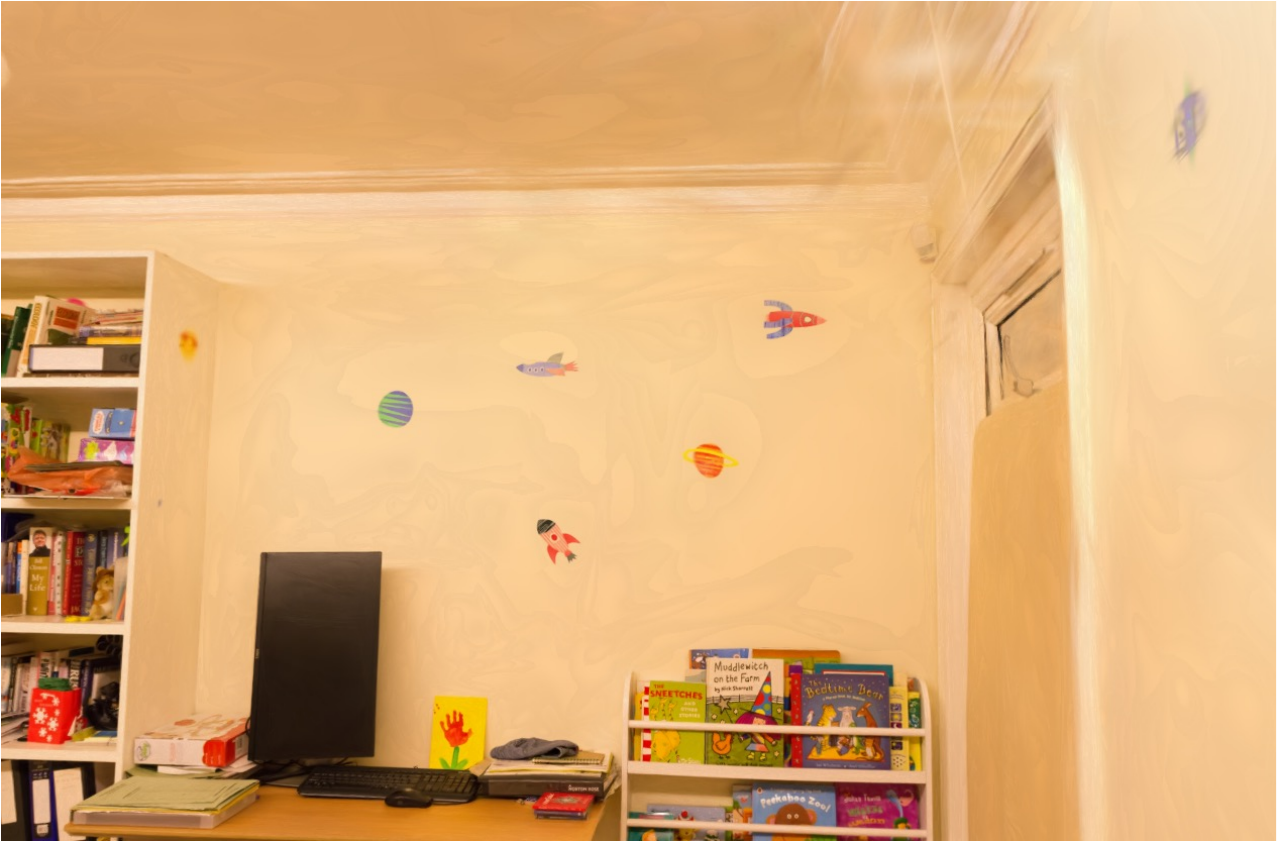} \\
      \cmidrule(lr){1-2} \cmidrule(lr){3-8}
      \multicolumn{2}{c}{\textbf{Type}} &
      \multicolumn{6}{c}{Real World (Indoor \& Outdoor)}\\
      \midrule
      \multicolumn{2}{c}{\textbf{Resolution}} & 980$\times$545 & 979$\times$546 & 1559$\times$1039 &
      1557$\times$1038 & 1332$\times$876 & 1264$\times$832 \\
      \midrule
      \multicolumn{2}{c}{\textbf{\# of Gaussians}} & 1.46 M & 2.43 M & 1.13 M & 0.76 M & 1.72 M & 0.97 M \\
      \midrule
      \multicolumn{2}{c}{\textbf{BVH Height (20-tri)}} & 27 & 28 & 26 & 26 & 26 & 27 \\
      \midrule
      \multirow{2}{*}{\textbf{BVH Size}} & \textbf{20-tri} & 2.34 GB & 3.88 GB & 1.81 GB & 1.21 GB & 2.75 GB & 1.54 GB \\
                                         & \textbf{TLAS+20-tri} & 208 MB & 345 MB & 161 MB & 107 MB & 245 MB & 137 MB \\
      \midrule
      \multirow{2}{*}{\textbf{BVH Memory Footprint}} & \textbf{20-tri} & 160 MB & 181 MB & 159 MB & 150 MB & 121 MB & 77 MB \\
                                  & \textbf{TLAS+20-tri} & 33 MB & 36 MB & 30 MB & 21 MB & 15 MB & 13 MB \\
      \bottomrule
    \end{tabular}
  }
\end{table*}


\myparagraph{Workloads.}
\tabref{workloads} presents the workloads used for our evaluation.
We select six widely used scenes from diverse
datasets~\cite{kna:par17,bar:mil22,hed:phi18}, encompassing both indoor and
outdoor real-world scenes with varying levels of complexity.
For each scene, we train the Gaussian model for 30K iterations using the
original \emph{ray tracing-based} training implementation from
3DGRT~\cite{moe:mir24}, which results in approximately 0.8M to 2.4M Gaussians per
scene.
To ensure tractable simulation time, we render all scenes at
128$\times$128 pixel resolution while preserving the same field of view
(FoV) as the original viewpoints.

To evaluate the end-to-end performance of \name{}, we implement a Gaussian ray
tracing renderer in Vulkan. This is because the original 3DGRT implementation
uses the NVIDIA OptiX framework, which is incompatible with our simulator that
exclusively supports Vulkan-based ray tracing programs.
We employ the same approach described in 3DGRT~\cite{moe:mir24} to gather the
next $k$ closest Gaussians during a single traversal and perform early
termination when possible.
In~\secref{analysis}, we discuss the details of our Vulkan implementation
and show that it achieves performance similar to the original OptiX implementation.

\putssec{perf}{Performance}

\begin{figure}[t]
  \centering
  \includegraphics[width=\columnwidth]{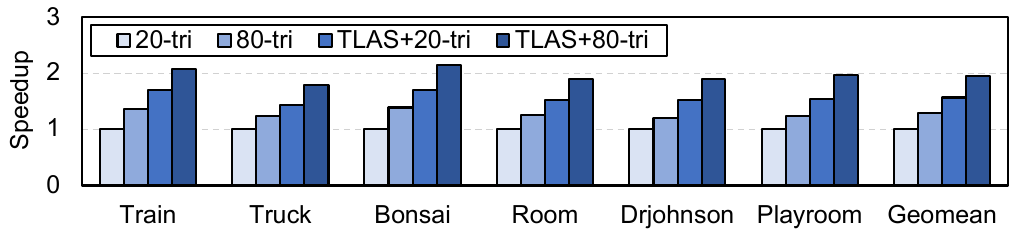}
  \caption{
    \name{}-SW performance with different Gaussian geometries.
  }
  \label{fig:grtx-sw-perf}
\end{figure}

\begin{figure}[t]
  \centering
  \includegraphics[width=\columnwidth]{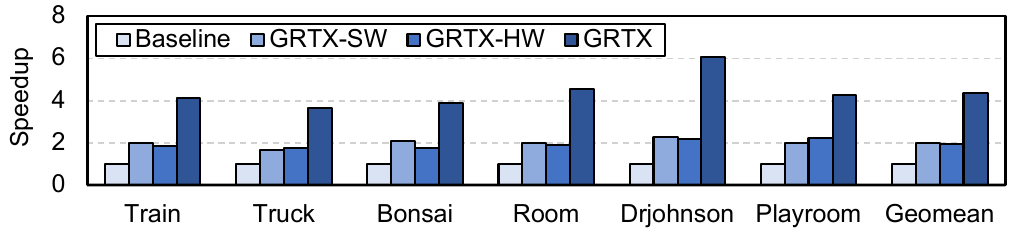}
  \caption{
    Speedup of \name{} over the baseline GPU using an icosahedron (i.e.,
    20-tri) as the bounding primitive.
  }
  \label{fig:performance}
\end{figure}


\myparagraph{Performance of \name{}-SW on Real GPU.}
\figref{grtx-sw-perf} shows the performance of \name{}-SW compared to the
monolithic BVH with \textit{20-faced}~\cite{moe:mir24} and
\textit{80-faced}~\cite{con:spe24} stretched polyhedrons on an RTX 5090 while
rendering images at 128$\times$128 resolution.
The results show that \name{}-SW provides noticeable speedups in both cases
through the optimization of acceleration structures.
Note that performance benefits may vary depending on rendering resolutions,
FoVs, or the case of second ray tracing, which we discuss in
Sections~\ref{ssec:sensitivity-study} and~\ref{sec:analysis}.


\myparagraph{End-to-End Performance of \name{}.}
\figref{performance} compares the end-to-end rendering performance of \name{}
against the baseline 3DGRT (20-tri) implementation.
\name{}-SW applies only the shared BLAS-based BVH construction (TLAS+20-tri)
without hardware modifications, whereas \name{}-HW adds only traversal
checkpointing to the baseline GPU. \name{} combines both optimizations.
Overall, \name{} achieves an average speedup of 4.36$\times$ (up to
6.09$\times$) over the baseline.

\name{}-HW avoids redundant node fetches and intersection tests across tracing
rounds, which results in a 1.94$\times$ speedup on average.
We observe that \name{}-HW delivers slightly higher speedups in scenes
containing large Gaussians (e.g., the walls in Drjohnson and Playroom).
In these cases, the large, overlapping bounding boxes of those Gaussians force
rays to traverse deeper into the BVH---even for Gaussians that ultimately
miss---thereby exacerbating redundant node visits across rounds.
Our checkpointing mechanism mitigates this by resuming traversal at lower-level
nodes, effectively bypassing redundant upper-hierarchy traversal.

We note that the benefits of \name{}-SW may be slightly higher in simulation,
as our infrastructure may not fully capture the characteristics of
state-of-the-art GPUs---though simulation results (2.00$\times$ average
speedup) reasonably align with real GPU behavior.
Nevertheless, with the results in~\figref{grtx-sw-perf}, we can conclude that
\name{} substantially improves the rendering performance of Gaussian ray
tracing by reducing the amount of traversal work and improving node fetch
locality.

\begin{figure}[t] 
  \centering
  \includegraphics[width=\columnwidth]{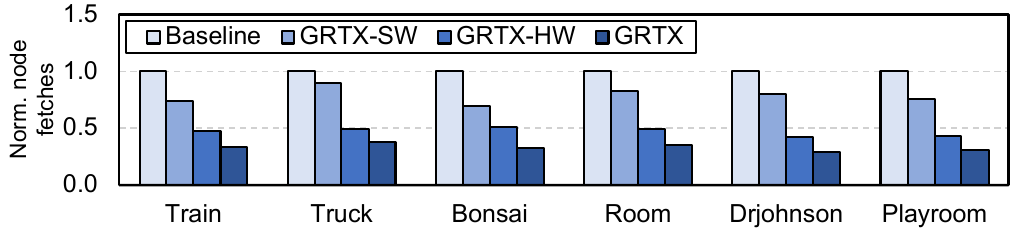} 
  \caption{Number of node fetches normalized to baseline.} 
  \vspace{-0.15in}
  \label{fig:num-node-fetches} 
\end{figure}

\putssec{src-of-gain}{Source of Performance Gain}

\myparagraph{Reduction in Node Fetches.}
\figref{num-node-fetches} shows the number of BVH node fetches normalized to
the baseline (20-tri).
The results indicate that \name{} reduces the number of node fetches by
3.03$\times$ on average compared to the baseline.

\name{}-SW increases the likelihood that different rays fetch the same node by
leveraging shared BLAS. These duplicate requests are merged into a single
operation, thereby reducing overall node fetches.
The magnitude of this reduction, however, varies depending on scene
characteristics.
In general, scenes with higher leaf-to-total node access ratios (e.g., Bonsai)
tend to benefit more from \name{}-SW, since BLAS-level optimizations have a
greater impact in such cases.
Conversely, scenes with lower ratios (e.g., Truck) likely see reduced benefits, as
overall traversal costs are more influenced by upper-level TLAS nodes.

On the other hand, the baseline RT unit lacks information from previous rounds,
resulting in redundant node fetches and intersection tests.
\name{}-HW addresses this by checkpointing the traversal state and reusing it
in the next round, thereby eliminating the need to re-traverse nodes already
visited in earlier rounds.
As a result, \name{}-HW reduces node fetches by an additional
2.37$\times$ on average on top of \name{}-SW.
Since BVH traversal is the primary bottleneck in Gaussian ray tracing, avoiding
these redundant operations leads to noticeable performance improvements.

\begin{figure}[t]
  \centering
  \includegraphics[width=\columnwidth]{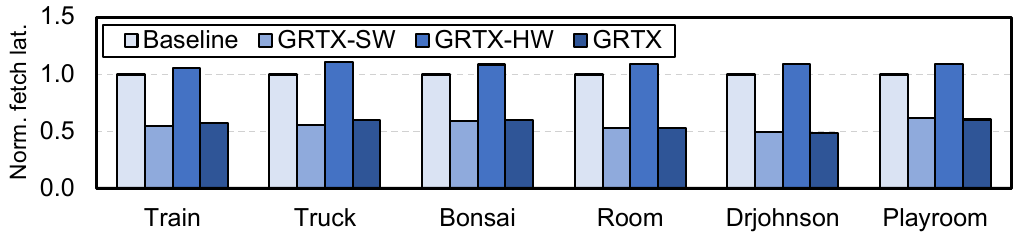}
  \caption{
    Average node fetch latency normalized to baseline.
  }
  \label{fig:avg-node-fetch-lat}
\end{figure}

\myparagraph{Node Fetch Latency.}
\figref{avg-node-fetch-lat} shows the average node fetch latency across
different configurations, all normalized to the baseline.
The baseline uses a monolithic BVH with a 20-triangle bounding mesh for each
Gaussian, resulting in a large BVH size. Consequently, many nodes are fetched
from lower levels of the memory hierarchy, leading to high fetch latency.
In contrast, \name{} employs a shared BLAS representing a unit sphere, which
offers two key advantages: its compact size allows it to fit entirely within
the L1 cache, and it enables high locality in node accesses during BLAS
instance traversal.

With the checkpointing and replay mechanism, rays resume traversal from
different nodes rather than uniformly starting from the root, which may reduce
initial ray coherence.
However, this does not result in a noticeable increase in average node fetch
latency because traversal paths quickly diverge regardless of the starting
point, making it difficult to exploit locality for node accesses even in the
baseline.
Overall, \name{} reduces the average node fetch latency by 1.77$\times$
compared to the baseline.

\begin{figure}[t]
  \centering
  \includegraphics[width=\columnwidth]{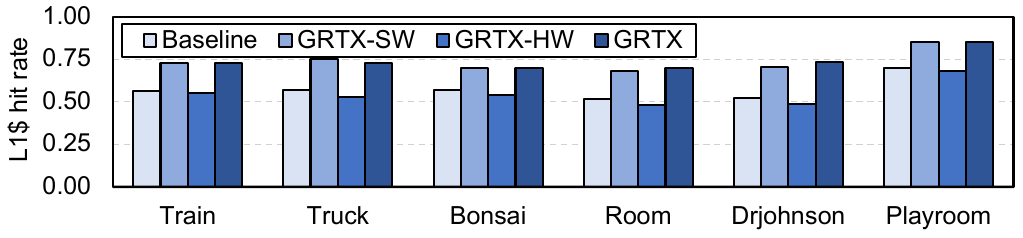}
  \caption{
    L1 cache hit rate for node fetches.
  }
  \label{fig:l1-hit-rate}
\end{figure}

\begin{figure}[t]
  \centering
  \includegraphics[width=\columnwidth]{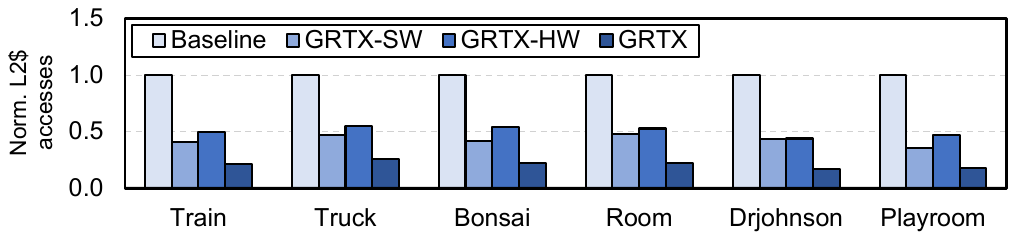}
  \caption{
    Total number of L2 cache accesses normalized to baseline.
  }
  \label{fig:l2-access}
\end{figure}

\myparagraph{L1 and L2 Cache Accesses.}
Figures~\ref{fig:l1-hit-rate} and~\ref{fig:l2-access} present the L1 cache hit
rate and the number of L2 cache accesses for the baseline (20-tri) and our
proposed approaches.
The baseline exhibits relatively low L1 cache hit rates across all scenes
due to the large memory footprint of its monolithic BVH structure.
In contrast, \name{}-SW achieves substantial improvements, with L1 cache hit
rates exceeding 70\% across all evaluated scenes.
This improvement stems from using a single shared BLAS for all Gaussian
primitives, which enhances temporal locality during BVH traversal and allows
the BLAS to reside within the L1 cache.
The higher L1 hit rate directly translates to lower node fetch latency, as more
BVH nodes are served from the fast L1 cache rather than from slower levels of
the memory hierarchy.

\name{} maintains nearly the same L1 cache hit rates as \name{}-SW while
further reducing L2 cache accesses.
The reduction primarily stems from eliminating redundant node fetches through
the checkpointing and replay mechanism. Overall, \name{} reduces L2 cache
accesses by 4.75$\times$ compared to the baseline.
These results demonstrate that the shared BLAS design and checkpointing
mechanism enable effective use of the cache hierarchy in Gaussian ray tracing,
thereby improving rendering performance.


\putssec{sensitivity-study}{Sensitivity Study}

\begin{figure}[t]
  \centering
  \includegraphics[width=\columnwidth]{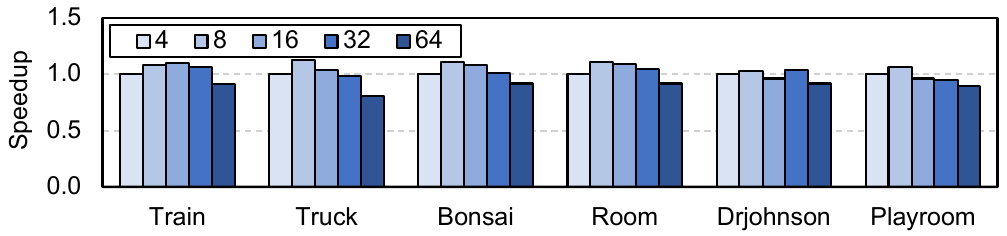}
  \caption{
    Performance comparison across different $k$-buffer sizes.
  }
  \label{fig:grtx-k-sweep}
\end{figure}

\myparagraph{$k$-Buffer Size.}
\figref{grtx-k-sweep} shows the performance of \name{} across different
$k$-buffer sizes, which is normalized to {$k=4$}.
Since \name{}-HW eliminates redundant node fetches and intersection tests in
subsequent traversal rounds, smaller $k$ values can reduce the total BVH
traversal cost by enabling more fine-grained ERT.
However, using smaller $k$ values leads to more traversal rounds (i.e.,
additional \texttt{traceRayEXT} calls), which in turn increases the overall
intra-warp synchronization overhead, as threads that complete traversal early
must wait for stragglers within the same warp to finish each round.
The results indicate that performance generally improves as $k$ decreases, but
the increased straggler overhead eventually offsets the traversal savings;
e.g., $k=4$ performs worse than $k=8$.
In our evaluation, we use $k=8$ as the default configuration since it
delivers the best average performance.

\begin{figure}[t]
  \centering
  \includegraphics[width=\columnwidth]{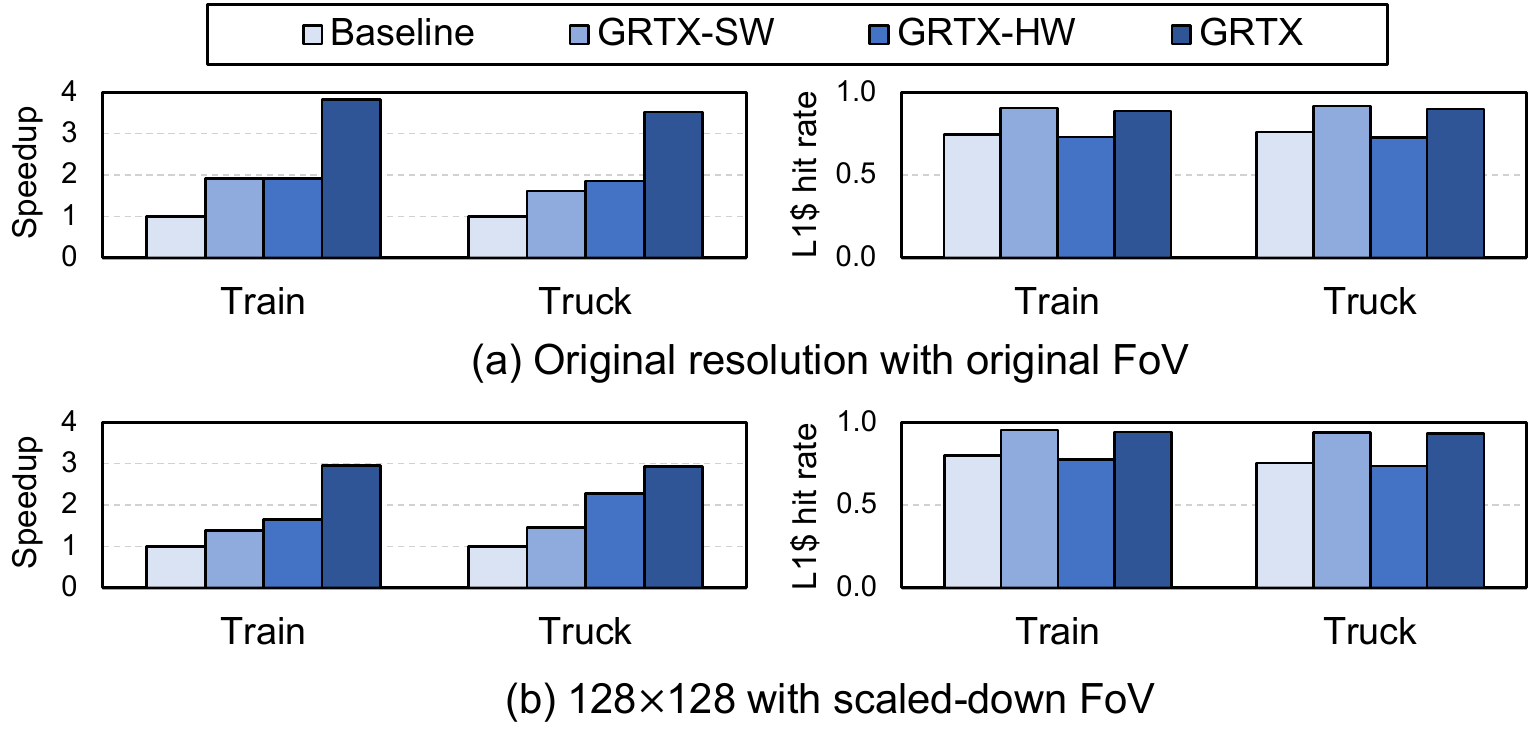}
  \caption{
    Performance and L1 cache hit rate of \name{} across different
    resolution and FoV settings compared to the baseline.
  }
  \label{fig:res-fov-sweep}
\end{figure}

\myparagraph{Varying Resolutions and FoVs.}
\figref{res-fov-sweep} presents the performance and L1 cache hit rate of
\name{} across different resolutions and FoVs.
\figref{res-fov-sweep}(a) evaluates the original resolutions (listed in
\tabref{workloads}) with the original FoVs, while \figref{res-fov-sweep}(b)
uses a 128$\times$128 resolution with proportionally scaled-down FoVs
(equivalent to cropping).
Higher resolutions and smaller FoVs both reduce the angular area per pixel,
thereby increasing ray coherence.
\name{}-HW provides consistent speedups across all scenarios, as it reduces
redundant per-ray BVH traversal, independent of ray coherence.
On the other hand, \name{}-SW exhibits lower relative speedups in
high-coherence scenarios, as coherent rays already achieve high baseline cache
locality.
Nevertheless, it still provides average speedups of {1.75$\times$} and
{1.43$\times$} for high-resolution and small FoV scenarios, respectively, by
reducing the memory footprint.

\putssec{}{Implementation Overhead}
\begin{table}[b]
  \centering
  \caption{Hardware cost.}
  \label{tab:hw-cost}
  \resizebox{\columnwidth}{!}{%
    \begin{tabular}{cc}
      \toprule
      \textbf{Hardware} & \textbf{Size} \\
      \midrule
      \multirow{3}{*}{Checkpoint buffer information}
      & (1-bit flag + 2B src offset + 2B dst offset) \\
      & $\times$ 32 threads/warp $\times$ 8 warps \\
      & + 8B src address + 8B dst address + 2B max size \\
      \midrule
      \textbf{Total} & \textbf{1.05 KB} \\
      \bottomrule
    \end{tabular}
    }
\end{table}

\begin{figure}[t]
  \centering
  \includegraphics[width=\columnwidth]{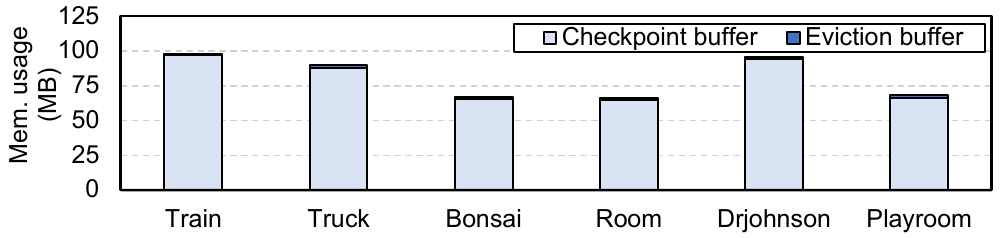}
  \caption{
    Memory usage of \name{} for checkpoint and eviction buffers.
  }
  \label{fig:mem-usage}
\end{figure}

\tabref{hw-cost} shows the additional storage required for checkpointing in the
warp buffer hardware.
Our hardware extensions require only 1.05~KB of storage per RT core.
Note that the checkpoint and eviction buffers used in \name{}-HW are allocated
in global memory, as discussed in~\ssecref{grtx-hw}, with their sizes bounded
by the maximum number of warps per SM multiplied by the number of SMs.
\figref{mem-usage} shows the memory usage of these buffers for our baseline
configuration (8 SMs).
For the Train scene, which exhibits the highest memory consumption, these
buffers consume only 97.68~MB combined.
Even when scaling to larger GPU configurations such as RTX 5090 (170 SMs), this
increases proportionally to 2.03~GB---just 6.3\% of the total 32~GB of GPU
memory.

\putsec{analysis}{Analysis and Discussion}

\begin{figure}[t]
  \centering
  \includegraphics[width=\columnwidth]{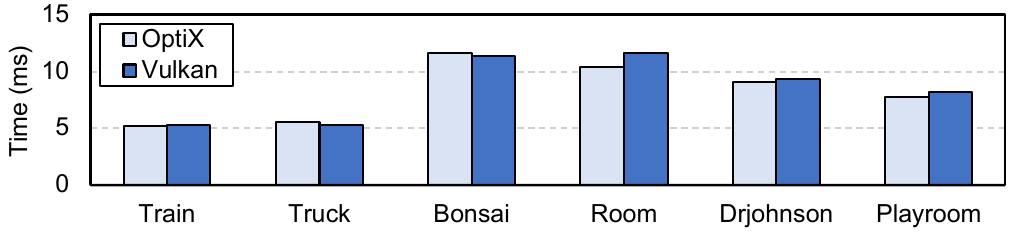}
  \caption{Rendering performance of the original OptiX implementation 
  of 3DGRT~\cite{moe:mir24} and our Vulkan implementation.}
  \label{fig:vulkan-perf}
\end{figure}

\myparagraph{Vulkan Implementation of 3DGRT.}
As mentioned in \ssecref{method}, we newly implement a Gaussian ray tracer in
Vulkan~\cite{vulkan} to run 3DGRT within our simulation framework, Vulkan-Sim.
Following the original 3DGRT~\cite{moe:mir24} implementation, we gather the
next $k$ closest Gaussians in the any-hit shader during a single traversal
round and perform blending and early ray termination in the raygen shader after
traversal.
The original 3DGRT implementation uses payload values to store all entries of
the $k$-buffer for each ray.
Since OptiX limits the maximum number of payload values to 32 and each
$k$-buffer entry requires two payload values, $k$ is fixed to 16 in the
original implementation.
Vulkan allows more flexible use of ray payloads, but we observe that allocating
the $k$-buffer within the payload structure results in a noticeable slowdown
compared to OptiX.
As such, we instead allocate the $k$-buffers of rays in global memory and
employ a Structure of Arrays (SoA) layout to make the memory access coalesced.

\figref{vulkan-perf} compares the rendering performance of the original OptiX
implementation and our Vulkan implementation.
We observe that our Vulkan implementation achieves performance similar to
OptiX.

\begin{figure}[t]
  \centering
  \includegraphics[width=\columnwidth]{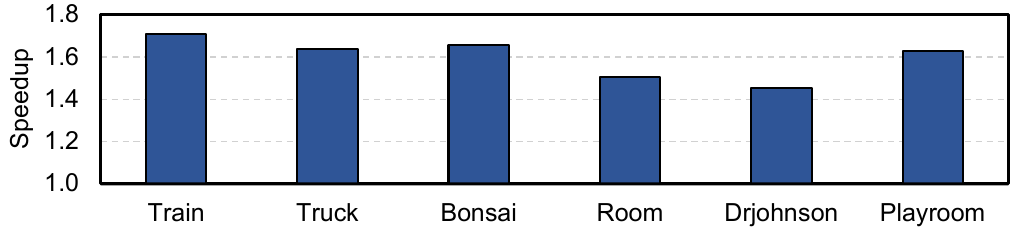}
  \caption{Speedup of \name{}-SW with sphere primitive over the baseline
  icosahedron mesh measured on RTX 5090.}
  \label{fig:grtx-sw-sphere-perf}
\end{figure}

\myparagraph{Using Sphere Primitive in \name{}-SW.}
In the NVIDIA Blackwell architecture, the RT core natively supports ray-sphere
intersection tests in hardware.
By exploiting this hardware support, we can implement \name{}-SW using a single
BLAS containing a unit sphere primitive, completely eliminating the need for
triangle meshes.
After transforming rays to Gaussian-local space, we only need one ray-box
and one ray-sphere intersection test per Gaussian, which is more efficient
than the triangle mesh-based approach.

\figref{grtx-sw-sphere-perf} shows the speedup of \name{}-SW with a sphere
primitive over the baseline icosahedron mesh, measured on the RTX 5090.
While the speedup is notable, we observe that the performance is lower than
TLAS+80-tri, as shown in \figref{grtx-sw-perf}, potentially due to the
throughput limitation of the ray-sphere intersection test in the current RT
core.
We expect that the performance will improve in future architectures with more
advanced RT cores that provide higher throughput for ray-sphere intersection
tests.

\begin{figure}[t]
  \centering
  \includegraphics[width=\columnwidth]{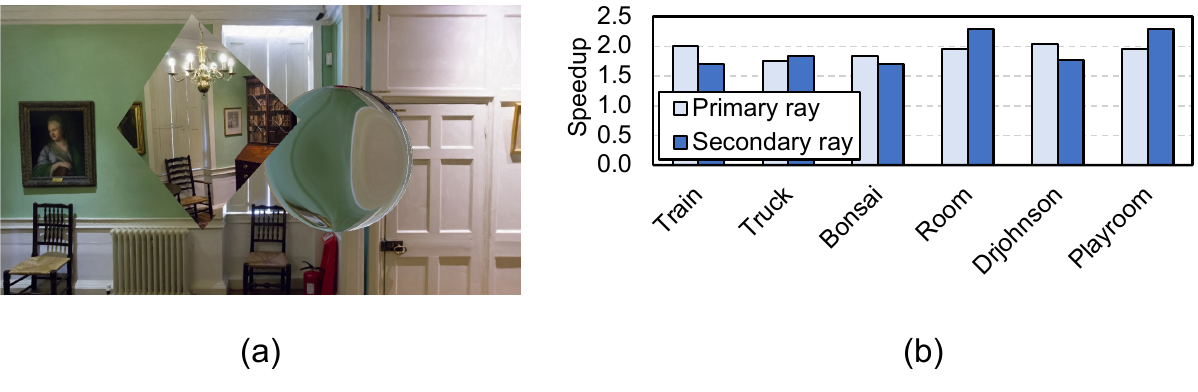}
  \caption{
    \name{}-HW performance on scenes with secondary ray effects. (a) An example
    image rendered with light effects (refractions and reflections). (b)
    Speedups for primary and secondary rays.
  }
  \label{fig:sec-rays}
  \vspace{-0.15in}
\end{figure}

\myparagraph{\name{}-HW on Secondary Rays.}
We evaluate the effectiveness of \name{}-HW on secondary rays.
To assess this, we augment each scene by adding a spherical glass object for
refractions and a rectangular mirror for reflections, both placed at random
locations, as illustrated in \figref{sec-rays}(a).
We then measure performance separately for primary rays (i.e., those cast from
the camera) and secondary rays (i.e., those generated by reflections and
refractions) to isolate the performance impact on each ray type.

\figref{sec-rays}(b) shows that \name{}-HW achieves similar speedups over the
baseline for both primary and secondary rays. This is because our checkpointing
mechanism reduces redundant traversal operations \emph{within} individual rays
rather than relying on ray coherence between different rays. 
Since replayed rays follow the exact same traversal paths as in previous rounds
regardless of ray type, incoherent secondary rays also benefit from \name{}-HW.

\myparagraph{Cross-Vendor Applicability.}
Modern RT accelerators vary in their implementations: some perform end-to-end
traversal, including intersection tests and node fetches (e.g., NVIDIA, Intel),
while others target only intersection operations with shader cores handling
node fetches (e.g., AMD).
Nevertheless, \name{} can provide performance benefits across GPU vendors as it
addresses fundamental traversal inefficiencies---redundant traversal,
divergence, and excessive memory footprint---that persist across all
architectures.

\begin{figure}[t]
  \centering
  \includegraphics[width=\columnwidth]{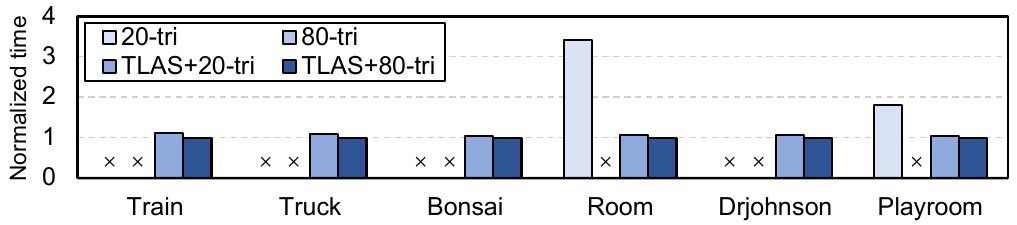}
  \caption{
    Normalized rendering time of baseline RT with monolithic BVH and
    \name{}-SW on an AMD GPU (Radeon RX 9070 XT). $\times$ indicates cases
    that cannot run as their BVHs exceed the maximum buffer allocation size
    (4 GB) in Vulkan.
  }
  \label{fig:amd-gpu-perf}
  \vspace{-0.10in}
\end{figure}

\figref{amd-gpu-perf} shows the rendering time of the baseline and \name{}-SW,
normalized to TLAS+80-tri, on the AMD Radeon RX 9070 XT.
We observe that AMD generates larger BVHs than NVIDIA, which causes baseline RT
with 20-/80-tri meshes to exceed the maximum buffer allocation size (4~GB)
in Vulkan for most scenes.
Our shared BLAS approach (TLAS+20-/80-tri) avoids this issue while achieving
1.73--3.42$\times$ speedup over the 20-tri baseline, demonstrating both
memory efficiency and performance benefits across vendors. 

\myparagraph{Support for Dynamic and Multi-Object Scenes.}
One might wonder whether our two-level BVH approach conflicts with traditional
dynamic scene rendering, which also uses two-level structures where each object
is a BLAS instance in a scene TLAS.
However, \name{} naturally extends to dynamic and multi-object scenes through
\emph{multi-level instancing} (supported in OptiX/HIP RT), creating a
three-level hierarchy: 1) a shared BLAS template for Gaussian primitives, 2)
per-object instances, and 3) a scene-level TLAS.
In this configuration, each Gaussian object maintains its own two-level
structure (\name{}-SW), while multiple objects are organized under the scene
TLAS for traditional dynamic scene management.
Object additions or removals require updating the scene TLAS, and object
movements require updating per-object transformation matrices---identical to
conventional dynamic rendering with no additional \name{}-specific overhead.

\putsec{related}{Related Work}

\myparagraph{Radiance Field Rendering Acceleration.}
Radiance field-based rendering, exemplified by Neural Radiance Fields
(NeRF)~\cite{mil:sri20}, has been actively studied for high-quality 3D scene
reconstruction and rendering. However, NeRFs suffer from slow training and
rendering, prompting numerous prior works to propose software
optimizations~\cite{mul:eva22,sun:sun22,fri:yu22,che:fun23} and hardware
accelerators~\cite{lee:cho23,li:li23,mub:kan23,son:wen23,fen:liu24,li:zha24,son:he24,noh:shi25}.
3D Gaussian Splatting~\cite{ker:kop23}, the current state-of-the-art method,
has attracted growing attention by achieving significantly faster rendering
than NeRFs through rasterization while maintaining high image quality.
Recent studies have also explored software and hardware optimizations to
further accelerate 3D Gaussian
Splatting~\cite{rad:ste24,he:li25,ye:fu25,lin:fen25,wu:zhu24,fen:lin25,pei:li25,wan:zhu25}.
Among these, GSCore~\cite{lee:lee24} and VR-Pipe~\cite{lee:kim25} are the first
to focus on hardware acceleration: GSCore proposes a dedicated accelerator,
while VR-Pipe introduces a novel extension to the hardware graphics pipeline.
However, both target rasterization-based Gaussian rendering.
In contrast, \name{} accelerates 3D Gaussian ray tracing by extending ray
tracing accelerators in modern GPUs. To our knowledge, \name{} is the first
work to analyze performance bottlenecks of 3D Gaussian ray tracing and propose
a hardware extension to existing GPU ray tracing accelerators.

\myparagraph{3D Gaussian Ray Tracing.}
To address the limitations of rasterization-based Gaussian rendering, several
studies~\cite{con:spe24,mai:hed25,bla:des25,yu:sat24}, including 3D Gaussian
Ray Tracing~\cite{moe:mir24} from NVIDIA, have demonstrated the potential of
ray tracing for Gaussian rendering using ray tracing accelerators.
While ray tracing hardware can effectively reduce rendering time, a significant
performance gap remains between rasterization and ray tracing. 
\name{} bridges this gap through optimized BVH construction for Gaussian
primitives and a minimal hardware extension to existing ray tracing
accelerators in modern GPUs.

\myparagraph{Ray Tracing Acceleration.}
Ray tracing architectures have been extensively explored in prior
work~\cite{nah:par11,sch:wal02,woo:sch05,nah:kwo14,lee:shi13}.
GPU vendors now integrate dedicated ray tracing accelerators, such as NVIDIA's
RT cores~\cite{rt-core}, into their GPUs.
Building on these, several studies have proposed techniques to further improve
ray tracing performance.
Ray predictor~\cite{liu:cha21} predicts ray intersections to skip traversal of
upper-level nodes in the acceleration structure.
While this effectively reduces traversal overhead when predictions are correct,
it is limited to ambient occlusion, which only requires detecting a
\emph{single} intersection.
However, 3D Gaussian ray tracing requires finding \emph{all} intersecting
Gaussians along the ray, making the ray predictor not directly applicable.
Treelet prefetching~\cite{cho:now23} prefetches nodes at treelet granularity to
reduce traversal latency in latency-bound ray tracing workloads. This technique
is orthogonal to our work, which focuses on reducing memory footprint and the
overall number of traversals.

\myparagraph{Accelerating General-Purpose Workloads with RT Units.}
There have been attempts to accelerate general-purpose workloads using
ray tracing units in GPUs~\cite{zhu22,ha:liu24,bar:she24,fen:li25,zha:zha25}.
RTNN~\cite{zhu22} leverages RT units for nearest-neighbor search via
software optimizations.
TTA~\cite{ha:liu24} and HSU~\cite{bar:she24} introduce hardware extensions that
enable traversal of general hierarchical data structures (e.g., trees) beyond
BVHs.
Heliostat~\cite{fen:li25} and RTSpMSpM~\cite{zha:zha25} extend RT units for
page table walks and sparse matrix multiplication, respectively.
In contrast, \name{} proposes software and hardware optimizations to accelerate
Gaussian ray tracing, an emerging and increasingly important application within
the conventional ray tracing domain.

\putsec{conclusion}{Conclusion}
Gaussian splatting has emerged as a leading technique for photorealistic image
synthesis, attracting widespread attention across academia and industry.
To overcome the inherent limitations of rasterization-based rendering, recent
work has explored rendering Gaussians via ray tracing.
However, existing methods suffer from low performance due to bloated
acceleration structures and redundant node traversals.
To address these inefficiencies, we introduce \name{}, which leverages
two-level acceleration structures and provides hardware support for
checkpointing and replay during BVH traversal.
With these software and hardware optimizations, \name{} greatly improves
the performance of Gaussian ray tracing over the existing method while
incurring minimal hardware overhead.

\section*{Acknowledgment}
We sincerely thank the anonymous reviewers for their valuable feedback.
This work was supported in part by the Institute for Information \&
Communications Technology Planning \& Evaluation (IITP) grants funded by the
Korean government (MSIT) (IITP-2026-RS-2022-00156295,
IITP-2026-RS-2023-00256081, IITP-2026-RS-2024-00395134).
The Institute of Engineering Research at Seoul National University provided
research facilities for this work. Jaewoong Sim is the corresponding author.


\bibliographystyle{IEEEtranS}
\bibliography{refs}

\end{document}